%% file: cvb_agn_in_clusters_astroph.tex
\documentclass[epsfig,useAMS,usenatbib]{mn2e}
\usepackage{epsfig,graphics,longtable}

\def\grtsim{\mathrel{\hbox{\rlap{\hbox{\lower2pt\hbox{$\sim$}}}\raise2pt\hbox{$>$}}}}
\def\lesssim{\mathrel{\hbox{\rlap{\hbox{\lower2pt\hbox{$\sim$}}}\raise2pt\hbox{$<$}}}}
\newcommand{\be}{\begin{equation}} \newcommand{\ee}{\end{equation}}
\title[A link between the evolution of clusters and their
  AGN fraction]{Evidence of a link between the evolution of clusters
  and their AGN fraction} \author[C. van Breukelen et al.]{Caroline
  van Breukelen,$^{1,2}$\thanks{Email: cvb@star.ucl.ac.uk} Chris
  Simpson,$^{3}$ Steve Rawlings,$^{2}$ Masayuki Akiyama,$^{4}$
  \newauthor David Bonfield,$^{5}$ Lee Clewley,$^{2}$ Matt J. 
  Jarvis,$^{6}$ Tom Mauch,$^{2}$ Tony Readhead,$^{7}$ \newauthor
  Ann-Marie Stobbart,$^{8}$ Mark Swinbank,$^{9}$ Mike Watson$^{8}$
  \\ $^{1}$Physics \& Astronomy, University College London, Gower
  Street, London, WC1E 6BT, UK\\ $^{2}$Astrophysics, Department of
  Physics, Keble Road, Oxford, OX1 3RH, UK\\ $^{3}$Astrophysics
  Research Institute, Liverpool John Moores University, Twelve Quays
  House, Egerton Wharf, Birkenhead CH41 1LD,UK\\ $^{4}$Subaru
  Telescope, National Astronomical Observatory of Japan, 650 North
  A'ohoku Place, Hilo, 96720 USA\\ $^{5}$NASA's Goddard Space Flight
  Center, Greenbelt, MD 20771, USA\\ $^{6}$Centre for Astrophysics,
  Science \& Technology Research Institute, University of
  Hertfordshire, Hatfield, AL10 9AB, UK \\ $^{7}$Owens Valley Radio
  Observatory, California Institute of Technology, Pasadena, CA 91125,
  USA\\ $^{8}$X-ray Astronomy Group, Department of Physics and
  Astronomy, University of Leicester, Leicester LE1 7RH,
  UK\\ $^{9}$Institute for Computational Cosmology, Department of
  Physics, Durham University, Durham DH1 3LE, UK\\} \date{Released
  2007 Xxxxx XX}

\pagerange{\pageref{firstpage}--\pageref{lastpage}} \pubyear{2002}

\def\LaTeX{L\kern-.36em\raise.3ex\hbox{a}\kern-.15em
    T\kern-.1667em\lower.7ex\hbox{E}\kern-.125emX}

\def\oii{[O\,\textsc{ii}]}
\def\oiii{[O\,\textsc{iii}]}
\def\as{^{\prime\prime}}
\def\am{^{\prime}}
\def\deg{^{\circ}}
\def\hr{^{\rm h}}
\def\mn{^{\rm m}}
\def\sc{^{\rm s}}

\begin{document}
\maketitle
\begin{abstract}
We discuss the optical properties, X-ray detections, and Active
Galactic Nucleus (AGN) populations of four clusters at $z \sim 1$ in
the Subaru-XMM Deep Field (SXDF). The velocity distribution and
plausible extended X-ray detections are examined, as well as the
number of X-ray point sources and radio sources associated with the
clusters. We find that the two clusters that appear virialised and
have an extended X-ray detection contain few, if any, AGN, whereas the
two pre-virialised clusters have a large AGN population. This
constitutes evidence that the AGN fraction in clusters is linked to
the clusters' evolutionary stage. The number of X-ray AGN in the
pre-virialised clusters is consistent with an overdensity of factor
$\sim 200$; the radio AGN appear to be clustered with a factor of
three to six higher. The median $K$-band luminosities of $L_K =
1.7~\pm~0.7~L^*$ for the X-ray sources and $L_K = 2.3~\pm~0.1~L^*$ for
the radio sources support the theory that these AGN are triggered by
galaxy interaction and merging events in sub-groups with low internal
velocity distributions, which make up the cluster environment in a
pre-virialisation evolutionary stage.
\end{abstract}


\section{Introduction}
A long-standing question in current astronomy is the connection
between the formation of large-scale structure and galaxy formation
and evolution. Studying clusters up to high redshifts gives us the
ideal opportunity to study the interaction between galaxies and the
intergalactic medium in detail, as a cluster's deep potential well
causes the cluster gas to be retained in the same
environment. Frequently studied phenomena impacting galaxy evolution
in clusters involve feedback mechanisms which couple the large-scale
gaseous environment to the small-scale generation of jets from
AGN. Jets and other AGN-driven outflows can heat and re-distribute the
gas, perhaps suppressing star-formation in the cluster
(e.g. Scannapieco \& Oh 2004; Fabian,Celotti \& Erlund 2006). As
emphasised by Rawlings \& Jarvis (2004), powerful radio jets can also have
profound influence on the evolution of galaxies in protoclusters. 

The correlation of radio-loud AGN and galaxy clusters has been studied
over a range of redshifts. At low redshift, luminous radio galaxies
tend to occur mostly in galaxy groups and low-mass clusters
(e.g. Prestage \& Peacock 1988, Hill \& Lilly 1991, Miller et
al. 2003). At higher redshifts ($z \sim 0.5$) however, it has been
shown that approximately 40\% of radio galaxies are located in massive
clusters of Abell richness 0 and higher (e.g. Hill \& Lilly
1991). Reaching a redshift of unity, some powerful radio sources are
found at the centres of galaxy clusters (e.g. Best 2000). Searches for
emission-line galaxies around radio galaxies at redshifts $z > 2$ have
shown that the latter often occur in (proto-) clusters (e.g. Venemans
et al. 2002)

The launch of the {\it Chandra X-ray Observatory} in 1999 made it
feasible to efficiently identify the prevalence of X-ray luminous AGN
in clusters. Optical follow-up of X-ray point sources in the fields of
rich clusters of galaxies have shown that clusters may contain a large
fraction of optically obscured AGNs (e.g. Martini et al. 2002,
2006). Martini, Mulchaey \& Kelson (2007) find that the fraction of X-ray selected
AGN is similar in clusters and the field, contrary to optically
selected AGN, although the fraction varies significantly between
clusters.

The distribution of AGN in clusters provides meaningful information on
the mechanism that triggers and sustains them. One such possible process
is the interaction and merging of galaxies (e.g. Barnes \& Hernquist
1996), enabling the creation of a central supermassive black hole and
the matter to fuel it. In this case the AGN fraction would be
determined by the properties of the environment providing the
opportunities for interaction and the supply of fuel. These external
conditions are likely to change in clusters as they evolve from
merging sub-groups to a massive virialised cluster. Studying clusters
at high redshifts and early evolutionary stages can therefore play a
key role in understanding the correlation between the AGN fraction and
their cluster environment.

In this paper we explore the AGN population of the highest-redshift
clusters found by Van Breukelen et al. (2006, henceforth VB06) in the
SXDF, using both radio and X-ray data to identify the AGN, and multi-object
spectroscopy on both the clusters and active galaxies to determine
their exact redshifts. This paper is organised as follows: in Section
2 we describe the spectroscopic observations and data
reduction; Section 3 presents the properties of each of the clusters
in our highest-redshift sample and in Section 4 we study the AGN
population of our clusters. Section 5 contains a discussion of our
conclusions. Throughout this paper we use the cosmological parameters
$H_0 = 71~\rm km~s^{-1}$, $\Omega_M = 0.3$ and $\Omega_{\Lambda} =
0.7$.

\section{The Data}

\subsection{The cluster sample}
\begin{figure}
\begin{center}
\includegraphics[height=7cm]{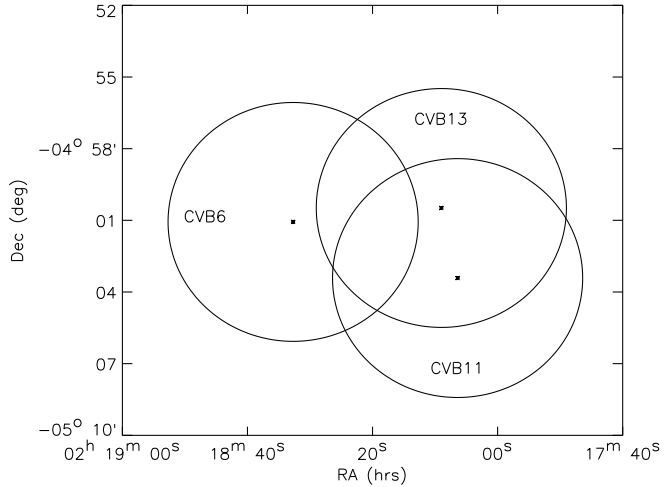}
\end{center}
\vspace{-0.2cm}
\caption{\small The positions of the three 5$\am$-radius cluster
  fields. CVB6 is centred on RA = $02\hr 18\mn 32.7\sc $, Dec = $-05\deg
  01\am 04 \as$; CVB11 is centred on RA = $02\hr 18\mn 06.4\sc$, Dec = $-05\deg
  03\am 25 \as$; and CVB13 is centred on RA = $02\hr 18\mn 09.0\sc $, Dec = $-05\deg
  00\am 29 \as$. Three-colour images of each of the fields can be
  found in Figs.\ref{overlay6}, \ref{overlay11}, and \ref{overlay13}.}
\label{fields}
\end{figure}

In this paper, we focus on the 5$\am$-radius cluster fields of CVB6
($z=0.9$), CVB11 ($z=1.1$), and CVB13 ($z=1.3$). These are the
highest-redshift clusters of VB06 with a significant number of
spectroscopically confirmed cluster members ($\grtsim 10$). The
positions of the three fields are depicted in Fig.~\ref{fields}. 

The field of CVB13 has been studied extensively using DEIMOS
spectroscopy by Van Breukelen et al. (2007, henceforth VB07). They
show the field contains two clusters: CVB13A at $z=1.28$ and CVB13B at
$z=1.45$. In this paper we focus solely on CVB13A as we now have a
larger number of confirmed cluster members available for this cluster
(see Section 3.3).

As will be discussed in Section 3.2, the cluster field of CVB11 also
contains two clusters: CVB11A at $z=1.06$ and CVB11B at $z=1.09$. For
the purposes of this paper, both clusters are included in our final
sample which consequently comprises four clusters in three fields. Note
that when we use the denominations `CVB11' or `CVB13' we are referring
to the cluster {\it fields}, whereas the postfix `A' or `B' signifies
the {\it individual} clusters.

\subsection{Imaging Data}
We use multi-wavelength data stemming from several surveys and
datasets. The optical imaging data (mainly used to create three-colour
images) are from the Subaru Telescope and comprise the $BVRi\am z\am$
bands (Furusawa et al. 2008). Near-infrared $J$ and $K$ data were
taken from the Ultra Deep Survey on the United Kingdom InfraRed
Telescope (UKIRT) (Foucaud et al. 2007). Further, we use X-ray data
from the {\it XMM-Newton} satellite (Watson et al. 2004) and radio
data (Simpson et al. 2006; Ivison et al. 2007) from the A- and B-array
configurations of the Very Large Array (VLA).

\begin{figure*}
\begin{center}
\includegraphics[height=19.5cm,angle=90]{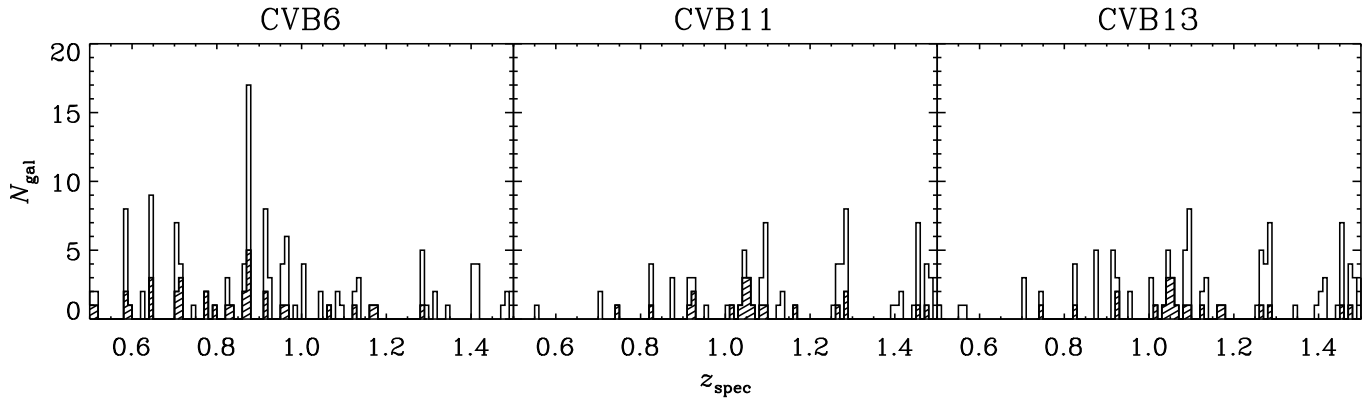}
\end{center}
\vspace{-0.5cm}
\caption{\small Spectroscopic redshift histogram for the three cluster
  fields at $0.5 \leq z_{\rm spec} \leq 1.5$. The shaded bins denote
  the galaxies observed with GMOS; the unfilled bins represent the
  DEIMOS targets. Note that the histograms for fields CVB11 and CVB13
  are very similar as they largely overlap.}
\label{spechist}
\vspace{-0.4cm}
\end{figure*}

\subsection{Spectroscopic Data}
The spectroscopic data used in this paper originate from various
sources\footnote{Note that due to the fact that the
spectroscopic data have been assembled from so many different sources,
we do not assume the samples of cluster galaxies to be
spectroscopically complete. However, we believe the selection function
is not biased to one particular type of object, as many different
galaxy populations have been targeted by the various spectroscopic
projects.}. Firstly, a subset of the cluster galaxies was observed with
the Deep Imaging Multi-Object Spectrograph (DEIMOS) on the Keck 2
telescope in Hawaii. The target selection, observations, and data
reduction can be found in VB07. Secondly, we make use of the SXDF
spectroscopic master list (maintained by C. Simpson and M. Akiyama,
private communication). This list contains redshifts for sources in
the SXDF/UDS derived from a number of observing runs on several
telescopes. The data we use in this paper result from VIsible
Multi-Object Spectrograph (Simpson et al. in preparation) and Faint
Object Camera and Spectrograph (FOCAS) (Yamada 2005 and Akiyama et
al. in preparation). Finally, the remainder of the cluster galaxies
were observed on the Gemini North telescope in Hawaii with the Gemini
Multi-Object Spectrograph (GMOS, Hook et al. 2004).  These data are
described below.

\subsubsection{GMOS target selection}
The targets for the spectroscopic data taken with GMOS were five
candidate clusters, identified in VB06. We selected them from the
clusters at photometric redshifts $0.8 \lesssim z_{\rm phot} \lesssim
1.0$. Since our telescope time was limited, we only targeted the
clusters that had a possible associated X-ray detection. The resulting
candidate clusters are given in the Appendix in Table~\ref{targets}.

For each cluster candidate, the target cluster galaxies for the MOS
mask were selected based on the cluster-selection algorithm outlined
in VB06. The algorithm presented in this paper used two methods to
detect clusters: Voronoi Tesselations and Friends-of-Friends. To
optimise the mask design, we divided the target galaxies into three
priorities for each cluster:
\begin{enumerate}
\item[{\bf Priority 1:}] all galaxies that were assigned to the
  cluster by both methods of the algorithm of VB06.
\item[{\bf Priority 2:}] the galaxies that were assigned to the
  cluster by either of the methods of the algorithm of VB06.
\item[{\bf Priority 3:}] all galaxies in the field-of-view of the GMOS
  instrument ($5.5\am \times 5.5\am$) with a photometric redshift
  within a 2-sigma range of the photometric redshift of the cluster
  candidate (see Table~\ref{targets} in the Appendix).
\end{enumerate}

The number of target galaxies for each cluster and the number of
galaxies included in the MOS masks are given in Table~\ref{targets} in
the Appendix.

\subsubsection{Spectroscopy on Gemini}
The data from GMOS were taken between 2006 August 17 and December 25
in queue mode (program ID: GN-2006B-Q-44). The MOS mask contained
slitlets of 1$\as$ wide and 3$\as$ long. To optimise the sky
subtraction during data reduction, we used the Nod \& Shuffle (N\&S)
mode with micro-shuffling. Each of our science exposures was divided
into 28 N\&S cycles of 60 seconds each, with 1.5$\as$ nodding offsets
on the sky and 3$\as$ shuffling offsets on the CCD. We used the R400
grating with no filter and a central wavelength of 795\,nm. The
spectral resolution of this set-up was $\lambda/\Delta\lambda \approx
1700$. Each target cluster was observed in four integrations of 3360
seconds. To reduce the effect of charge traps, cosmic rays, bad
pixels, and the gap between the two GMOS CCDs, each integration was
offset by 5\,nm in central wavelength (x-direction on the CCD) and a
DTA-X offset was introduced of 0, 2, and 4 pixels (y-direction on the
CCD). The binning on the CCD was $2 \times 2$ pixels in the spatial
and spectral directions, with an unbinned pixel size of 0.07$\as$ per
pixel. We maximised the number of cluster galaxies that could be
observed in each MOS mask by choosing the optimal position angle of the
instrument, which is given for each target cluster in
Table~\ref{targets} in the Appendix. The seeing was $\lesssim 0.80\as$
for all targets and conditions were photometric throughout.

For calibration purposes, spectroscopic flatfields were taken after
each exposure with a quartz-halogen lamp. We also executed a series of
35 darks to enable the removal of charge-traps during data
reduction. Per target cluster, one arc exposure with a quartz-halogen
lamp was taken for each central wavelength set-up. Finally, to allow
flux calibration, we included observations of the spectrophotometric
standard star BD+28d4211, using a longslit of 1$\as$ width.

\subsubsection{GMOS data reduction}
The first step in the data reduction was the bias subtraction of the
science and calibration frames, using bias exposures of the
corresponding observing dates and set-ups taken from the Gemini
archive. The science frames were sky subtracted and mosaicked using
the Gemini data reduction tasks for IRAF. Next we combined the darks
using the median value and identified the charge traps. These were
subsequently masked out in the science frames. Finally the four
science frames per target cluster were combined using the average
value and a three-sigma clipping routine to remove cosmic rays and bad
pixels.

We performed the flat-fielding, rectifying, cleaning and wavelength
calibrations using a set of Python routines (Kelson, private
communication). The final two-dimensional science frames were obtained
by shifting the reduced image by the N\&S offset and subtracting it
from the original reduced image. The sensitivity function was derived
in IRAF from the reduced spectrum of the standard star; this allowed
the flux calibration of the two-dimensional science frames. We
extracted the one-dimensional spectra using a boxcar extraction
routine with an aperture of 1$\as$.

\subsection{Redshift determination}
To determine the approximate redshifts of the galaxies observed both
with GMOS and DEIMOS, we identified strong spectral features such as
the \oii$_{3727}$, H$_\beta$, and \oiii$_{4960,5008}$ emission lines,
the 4000-\AA\ break, the Ca H\&K absorption lines at 3933.4 and
3969.2\,\AA\ and the $G$ band at 4304.4\,\AA. For all galaxies showing
the \oii~emission line we determined the exact redshift by fitting a
double Gaussian profile to the observed line profile, where the Full
Width Half Maximum (FWHM) of each Gaussian was assumed to be equal,
and was a free parameter of the fitted function. The other parameters
were the redshift, the continuum level, and the ratio of fluxes of the
two lines (see also VB07). The exact redshifts of the galaxies that
only show absorption features were measured by cross-correlating their
spectra with template spectral energy distributions. For this purpose
we used a set of three stellar population synthesis templates from
Bruzual \& Charlot (2003), consisting of solar metallicity, 1-Gyr
burst models of ages 3, 5, and 7 Gyr.

The spectroscopic redshifts obtained from the DEIMOS and GMOS data on
each of the cluster fields of our sample are shown in
Fig.~\ref{spechist}. In the Appendix we show the result of a
comparison between all spectroscopic redshifts and the photometric
redshifts determined in VB06. 

\begin{figure*}
\hspace{-1.5cm}
\includegraphics[width=19cm]{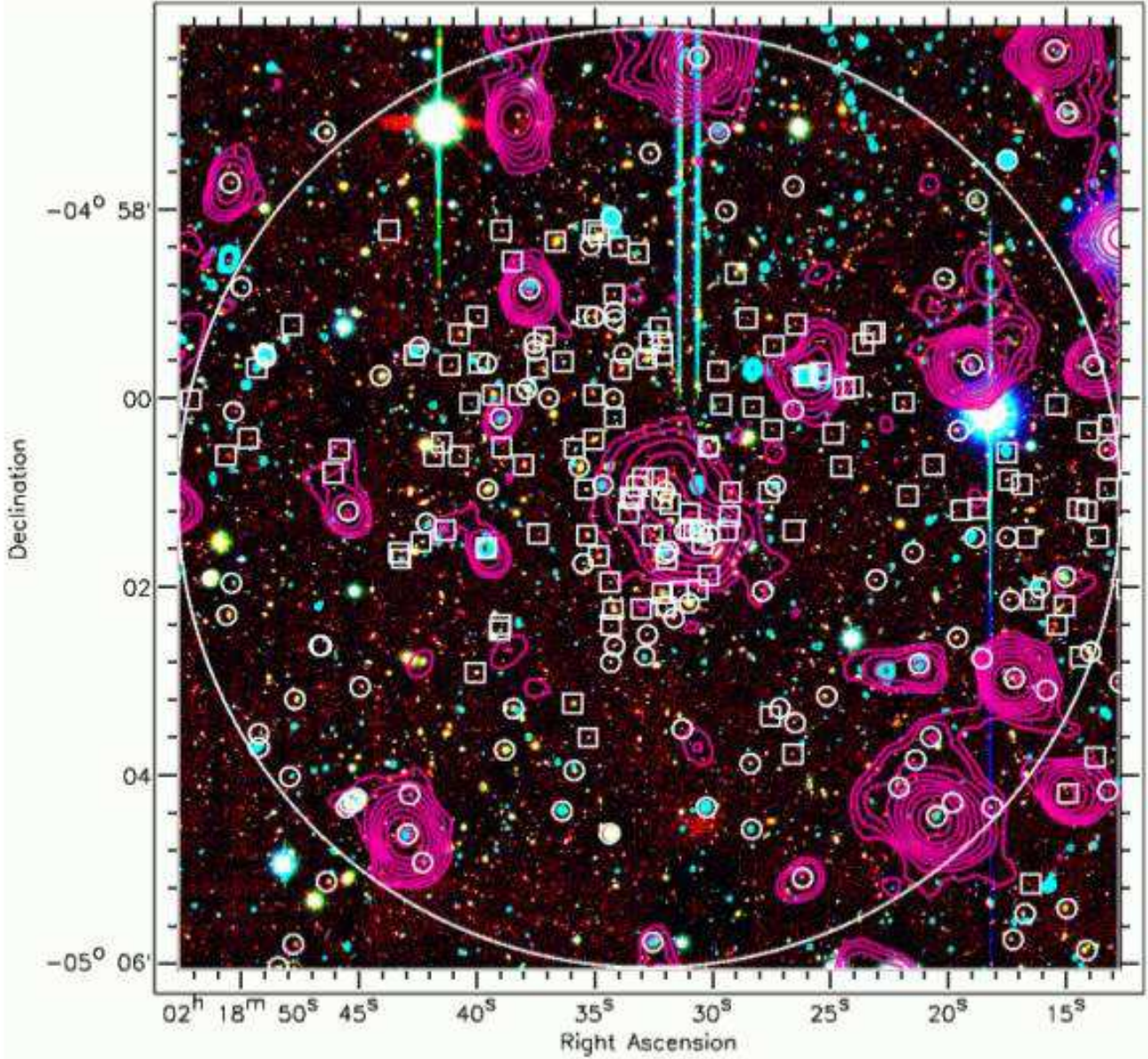}
\caption{\small $Bi^{\prime} K$ image of cluster field CVB6. The large
  circle encompasses the $5\am$-radius field we investigate in this
  paper. The spectroscopic targets are marked by the white symbols:
  squares denote objects observed with GMOS and DEIMOS and circles are
  from the SXDF master list. Broadband {\it XMM-Newton} X-ray contours
  from a signal-to-noise map are overlaid in purple and VLA A-array
  (with a beam size of 1.9$^{\prime\prime}$ by 1.6$^{\prime\prime}$ at
  PA = 22$^\circ$) contours in blue. The X-ray contours are at
  [$\sqrt{2}\sigma, 2\sigma, 2\sqrt{2}\sigma, ...$], and the radio
  contours at [$2\sqrt{2}\sigma,4\sigma, 4\sqrt{2}\sigma, ...$].}
\label{overlay6}
\vspace{-0.25cm}
\end{figure*}

\section{Cluster properties}\label{properties}

\subsection{CVB6}\label{cluster_cvb6}

\subsubsection{Optical properties}
The combined DEIMOS and GMOS spectroscopic data yielded 20 confirmed
cluster galaxies for CVB6. This is the largest spectroscopic dataset
we have available on any of our clusters. Fig.~\ref{overlay6} shows
all the data sets we use on the cluster field of CVB6: it is a
$Bi^{\prime} K$ image with spectroscopic targets marked and X-ray and
radio contours overlaid.

\begin{figure}
\includegraphics[width=85mm]{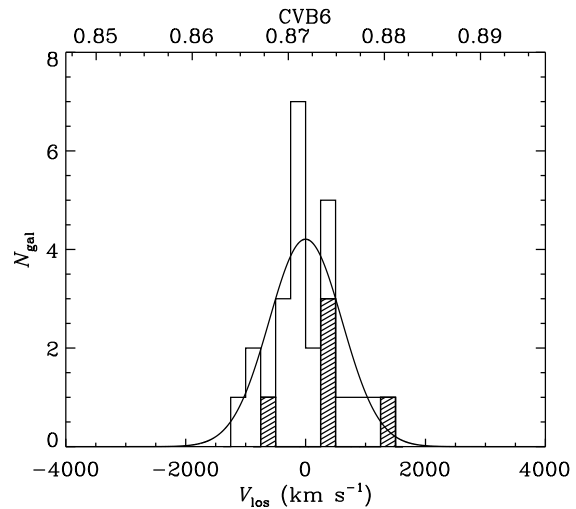}
\vspace{-0.75cm}
\caption{\small Velocity distribution of cluster CVB6. The overplotted
  Gaussian function is determined by $\overline z =
  0.87180\,\pm\,0.00007$ and $\sigma_v = 608\,\pm\,115\,\rm
  km\,s^{-1}$. The shaded bins are objects from the SXDF master list.}
\label{cvb6_veldisp}
\end{figure}

Table~\ref{cvb6} in the Appendix lists all cluster galaxies observed
with GMOS and DEIMOS with their properties. We calculate the cluster
redshift by taking the bi-weighted mean of the cluster galaxies
(including the objects from the SXDF master list) as outlined by Beers
et al. (1990). The velocity dispersion of the cluster is determined by
selecting all galaxies within $\pm\,2000\,\rm km\,s^{-1}$ of the
cluster redshift, and calculating the bi-weighted estimate of the
scale factor of the distribution, which is assumed to be
Gaussian. Fig.~\ref{cvb6_veldisp} shows the velocity distribution of
the cluster, and the associated Gaussian function determined by
$\overline z = 0.87180\,\pm\,0.00007$ and $\sigma_v =
608\,\pm\,115\,\rm km\,s^{-1}$. To calculate the virial mass of CVB6
we use the following empirical relation found by Evrard et al. (2008)
through N-body simulations: \be M_{200} = \frac{10^{15}h^{-1}{\rm
    M_{\odot}}}{H/H_0}\Big(\frac{\sigma_{\rm v,los}}{1083\,{\rm
    km\,s^{-1}}}\Big)^3,
\label{mass}
\ee where $M_{200}$ is the mass contained within a sphere of radius
$r_{200}$ for which the mean density is $200$ times the critical
density $\rho_{\rm cr}$. We arrive at $M_{200} = 1.6 \times
10^{14}\,\rm M_{\odot}$ for CVB6.

\begin{figure}
\vspace{-0.5cm}
\begin{center}
\includegraphics[width=80mm]{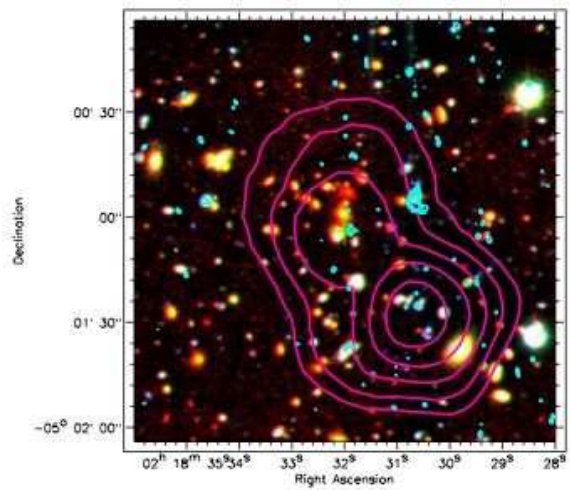}
\end{center}
\caption{\small $Bi^{\prime} K$ image of the central $2\am$ of cluster
  CVB6 with broadband X-ray contours (from a signal-to-noise map) and
  1.4\,GHz radio contours overlaid in purple and blue
  respectively. The green cross denotes the position of the extended
  X-ray source. The X-ray point source to the southwest is a
  background object at $z=3.0$. The optical data is from the Subaru
  Telescope, the infrared data from UKIRT, the X-ray data from {\it
    XMM-Newton}, and the radio data from the VLA (A-array with a beam
  size of 1.9$^{\prime\prime}$ by 1.6$^{\prime\prime}$ at PA =
  22$^\circ$). The X-ray contours are at [$\sqrt{2}\sigma, 2\sigma,
    2\sqrt{2}\sigma, ...$], and the radio contours at
  [$2\sqrt{2}\sigma,4\sigma, 4\sqrt{2}\sigma, ...$].}
\label{zoom_overlay6}
\end{figure}

\subsubsection{X-ray emission from the intracluster medium}

The 2XMM source catalogue (Watson et al. 2008) contains an X-ray source
coincident with the position of cluster CVB6 which is extended over
20.4$\as$ (the XMM point spread function is œôøÜ
6$\as$). Fig.~\ref{zoom_overlay6} shows a three colour image of the
central $2\am$ of CVB6 with X-ray and radio contours overlaid. The
extended emission is evident in the centre; to the southwest there is
a background X-ray point source of total flux $4.1\,\pm\,0.6 \times
10^{-17} \rm\,W\,m^{-2}$, associated with a spectroscopically
confirmed quasar at $z=3.0$.

\begin{figure*}
\begin{center}
\includegraphics[width=150mm, height=70mm]{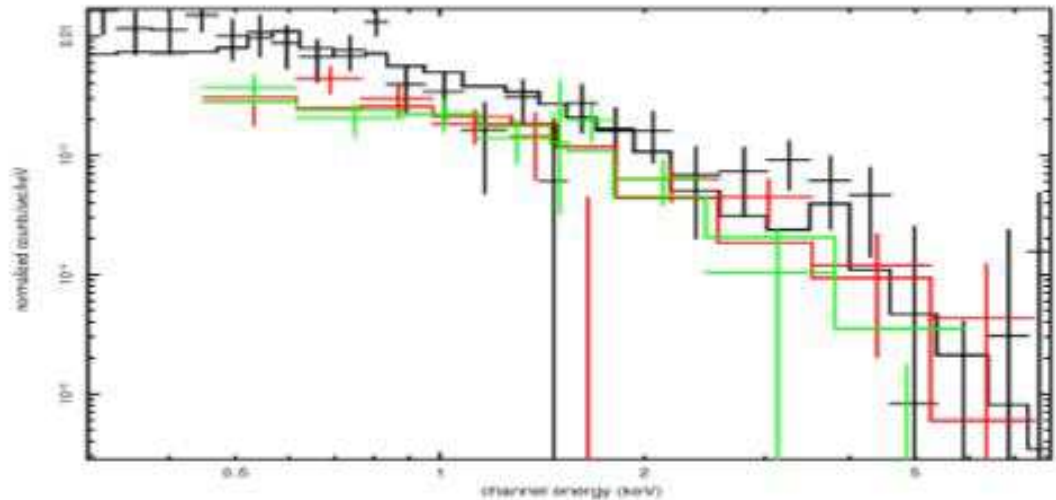}
\end{center}
\caption[X-ray spectrum of cluster CVB6]{\small X-ray spectrum of
  cluster CVB6. The green and red symbols are the data from the two
  individual X-ray cameras (green: PN, red: M). The black data points
  is the combined result. The black line is a fit to the data using a
  black body `mekal' emission spectrum and the `wabs' local absorption
  model.\bigskip}
\label{xrayspec}
\end{figure*}

The total X-ray flux of the extended source is $3.91\,\pm\,1.05 \times
10^{-17} \rm\,W\,m^{-2}$; its hardness ratios (HR) are $HR1 =
0.29\,\pm\,0.09$, $HR2 = -0.02\,\pm\,0.08$, $HR3 = -0.57\,\pm\,0.12$,
and $HR4 = 0.26\,\pm\,0.24$. Here the hardness ratios are defined as
$HR_i = (C_{i+1} - C_i) / (C_i + C_{i+1})$, where $C_i$ is the count
rate in band $i$. The energy bands are 1: 0.2 - 0.5\,keV, 2: 0.5 -
1.0\,keV, 3: 1.0 - 2.0\,keV, 4: 2.0 - 4.5\,keV, and 5: 4.5 -
12.0\,keV. The source is sufficiently bright to construct an X-ray
spectrum, which is shown in Fig.~\ref{xrayspec}. 
The X-ray luminosity at $z=0.87$ is calculated to be $L_{\rm X} = 1.45
\times 10^{37} \rm\,W$.

\begin{figure*}
\hspace{-1.2cm}
\includegraphics[width=170mm]{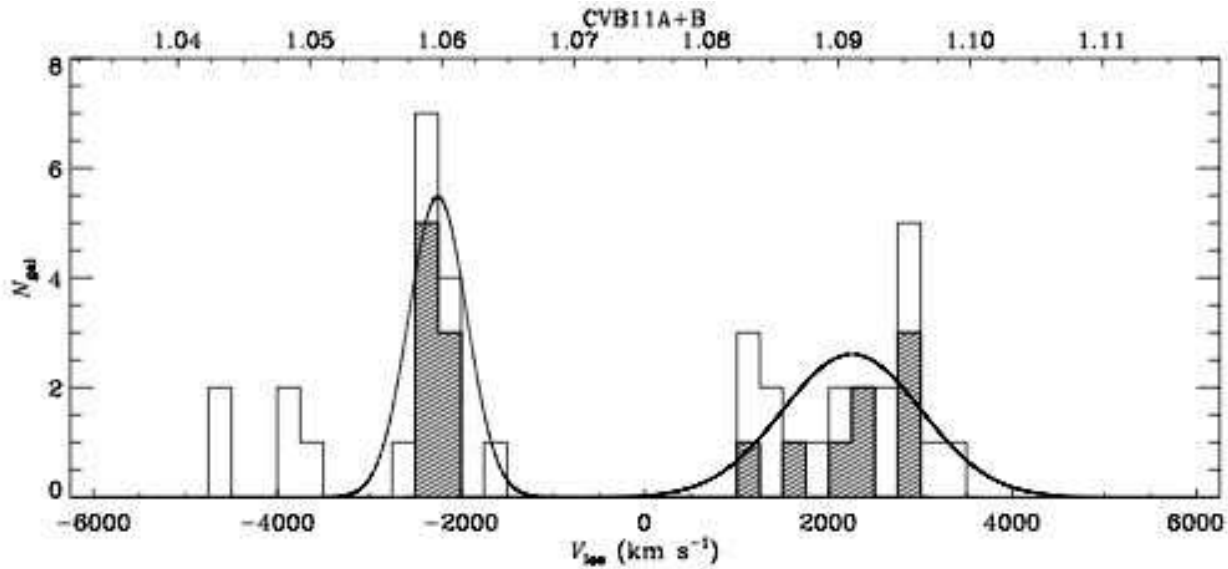}
\caption{\small Velocity distributions of clusters CVB11A and
  CVB11B. The overplotted Gaussian functions are determined by
  $\overline z = 1.0593\,\pm\,0.0003$ and $\sigma_v =
  316\,\pm\,166\,\rm km\,s^{-1}$ for CVB11A and $\overline z =
  1.091\,\pm\,0.001$ and $\sigma_v = 650\,\pm\,95\,\rm km\,s^{-1}$
  for CVB11B. The shaded bins are objects from the SXDF master
  list.\bigskip}
\label{cvb11_veldisp}
\vspace{-0.5cm}
\end{figure*}

Using the publicly available software package
\textsc{XSPEC}\footnote{http://xspec.gsfc.nasa.gov/docs/xanadu/xspec/index.html}
we fit a model spectral energy distribution to the X-ray spectrum. The
model consists of two multiplied components: (i) the `mekal' emission
spectrum from diffuse hot gas based on the model calculations of Mewe
and Kaastra (Mewe, Gronenschild \& van den Oord, 1985; Mewe, Lemen \&
van den Oord, 1986; Kaastra, 1992) with iron emission line
calculations by Liedahl, Osterheld \& Goldstein (1995) and (ii) the
`wabs' photo-electric absorption model using Wisconsin cross-sections
(Morrison and McCammon, 1983). The emission spectrum is determined by
the temperature of the intracluster medium; the best-fitting value is
$kT_{\rm X} = 4 \,\pm\,1.1\,\rm keV$ (rest-frame). The X-ray
luminosity and velocity dispersion of CVB6 are exactly as expected
according to the X-ray scaling relations for groups and clusters found
by Xue \& Wu (2000); the temperature is slightly higher than average
but still within the scatter of the observed relations.

\subsection{CVB11}\label{cluster_cvb11}
\begin{figure*}
\hspace{-1.5cm}
\includegraphics[width=19cm]{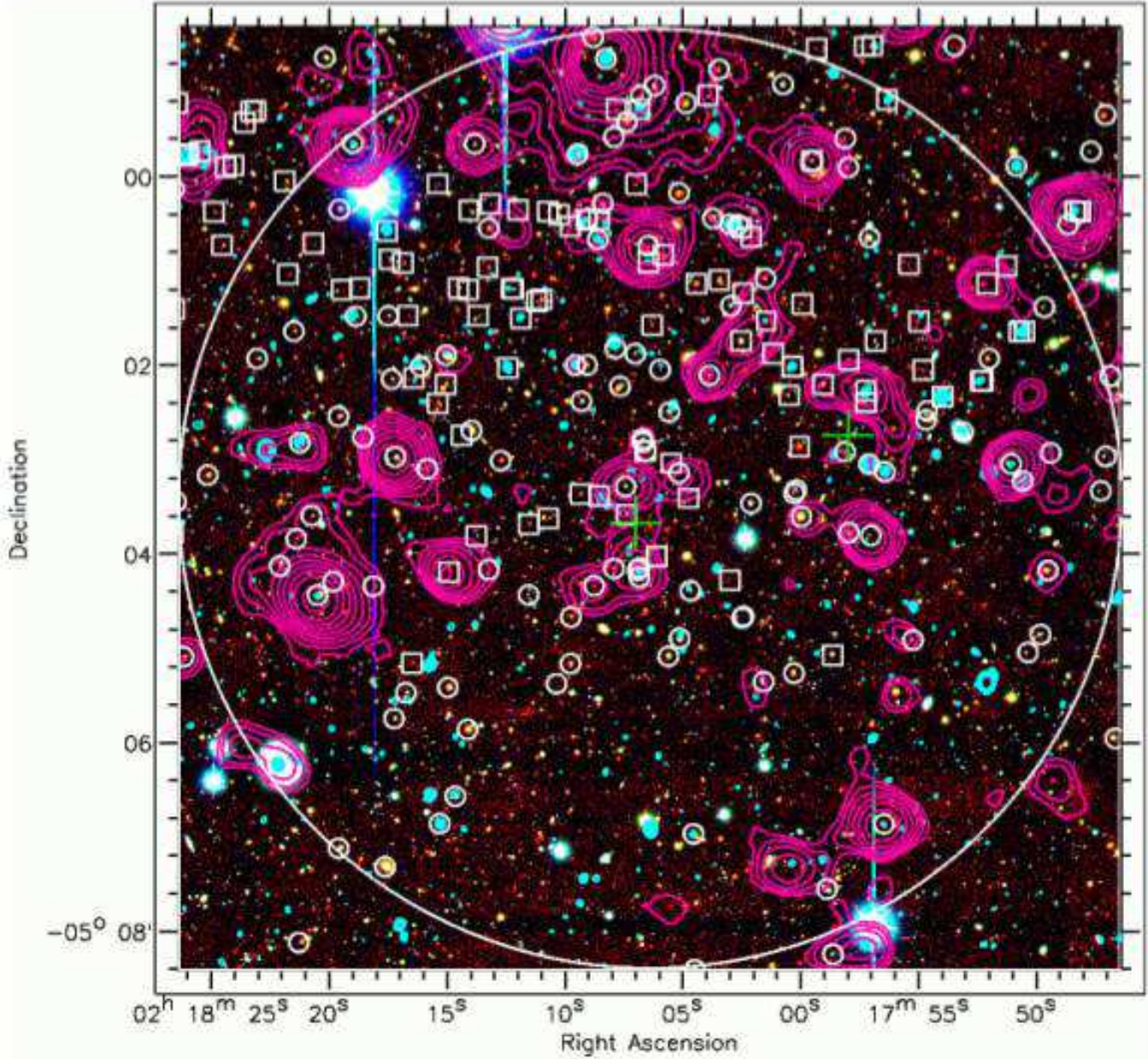}
\caption{\small $Bi^{\prime} K$ image of cluster field CVB11. The
  green crosses denote the approximate positions of clusters CVB11A
  and CVB11B, based on the averages of the positions of the respective
  cluster galaxies. CVB11A is in the centre and CVB11B to the
  northwest. The other symbols and contours are as in
  Fig.~\ref{overlay6}.}
\label{overlay6}
\vspace{-0.25cm}
\end{figure*}

\subsubsection{Optical properties}

Cluster CVB11 does not consist of a single redshift peak, but rather
comprises two peaks at $z = 1.06$ and $z = 1.09$, with $\Delta V \sim
4500\,\rm km\,s^{-1}$. In this paper we designate the peak at $z=1.06$
with CVB11A, and the peak at $z=1.09$ with CVB11B. CVB11A has 9
confirmed cluster galaxies in our GMOS and DEIMOS data, and CVB11B has
11 cluster members. All 20 observed cluster galaxies of CVB11A and
CVB11B are \oii~emitters. The properties of the cluster galaxies are
given in Table~\ref{cvb11} in the Appendix.

Fig.~\ref{cvb11_veldisp} shows the velocity distributions of CVB11A
and CVB11B, including the galaxies from the SXDF master list. The
exact cluster redshift and velocity dispersions are calculated by
using the bi-weighted mean of the galaxy redshifts and the `gapper'
estimate of the scale factor. Note that for CVB6 we used the
bi-weighted estimate for the scale factor; the appropriate estimator
needs to be chosen according to sample size, as discussed in Beers et
al. (1990). We arrive at $\overline z = 1.0593\,\pm\,0.0003$,
$\sigma_v = 316\,\pm\,166\,\rm km\,s^{-1}$ for CVB11A, and $\overline
z = 1.091\,\pm\,0.001$, $\sigma_v = 650\,\pm\,95\,\rm km~s^{-1}$ for
CVB11B. These velocity dispersions would, according to Eq. 1, relate
to masses of $2.0 \times 10^{13}$ and $1.7 \times 10^{14}
\rm\,M_{\odot}$ respectively. It is, however, unlikely that the latter
is a true estimate of the cluster mass as the velocity distribution of
cluster CVB11B from Fig.~\ref{cvb11_veldisp} does not appear to have
achieved the Gaussian distribution expected in line-of-sight
velocities of a virialised system. Unfortunately, the small sample
size complicates the calculation of reliable statistics on the
probability that the velocities are drawn from a Gaussion
distribution. Using the method of Marshall et al. 1983, we execute a
Bayesian likelihood test which is designed to choose the optimum model
(Gaussian or flat velocity distribution) given the data. There is
greater evidence for the flat model than the Gaussian model, however
owing to the sparseness of the data the probability of the data being
drawn from either distribution are only between 5 and 10\%, which
means neither can be confirmed or discarded reliably statistically.

\begin{figure}
\begin{center}
\includegraphics[width=90mm]{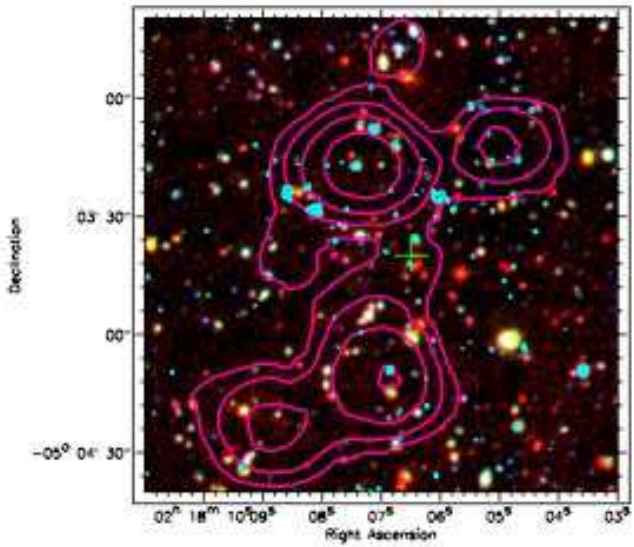}
\end{center}
\vspace{-0.5cm}
\caption{\small $Bi^{\prime} K$ image of the central $2\am$ of cluster
  CVB11A with X-ray and 1.4\,GHz radio contours overlaid in purple and
  blue respectively. The three brightest X-ray point sources are
  background objects at $z=1.4$ (to the north), $z=1.3$ (to the
  northwest) and $z=3.1$ (to the south). The emission in the centre
  between these sources is identified by Finoguenov et al. as an
  extended source (green cross), possibly associated with the
  cluster. Data sources and contour levels are as in
  Fig.~\ref{zoom_overlay6}. \bigskip}
\label{zoom_overlay11}
\end{figure}

\subsubsection{X-ray emission from the intracluster medium}
The X-ray catalogue does not contain any extended sources that could be
associated with cluster CVB11B. However, careful inspection of the
X-ray emission at the central position of CVB11A shows excess flux
between two bright point sources. Fig.~\ref{zoom_overlay11} shows a
three-colour image of the central $1\am$ of CVB11A with X-ray contours
overlaid in purple and radio contours overlaid in blue. The three brightest
X-ray point sources in this image are background sources at $z=1.3$,
$z=1.4$, and $z=3.1$ with total fluxes of $0.5 \,\pm\,0.2$, $0.5
\,\pm\,0.3$, and $2.4 \,\pm\,0.5 \times 10^{-17} \rm\,W\,m^{-2}$
respectively. Finoguenov et al. (in preparation) apply a sophisticated point
spread function removal technique to obtain fluxes for extended
sources which are polluted by point sources. They indeed find an
extended source at this position, with a flux of $1.9 \,\pm\,0.5
\times 10^{-18} \,\rm W\,m^{-2}$ in the 0.5 - 2.0 keV band. This would
mean a luminosity of $L_X = 1.2 \times 10^{36}\,\rm W$ if the X-ray
emission is associated with the cluster at $z=1.06$, which --
according to the scaling relations of Xue \& Wu (2000) --
corresponds well to the estimated velocity dispersion.

\subsection{CVB13}\label{cluster_cvb13}
\begin{figure*}
\includegraphics[width=18.5cm]{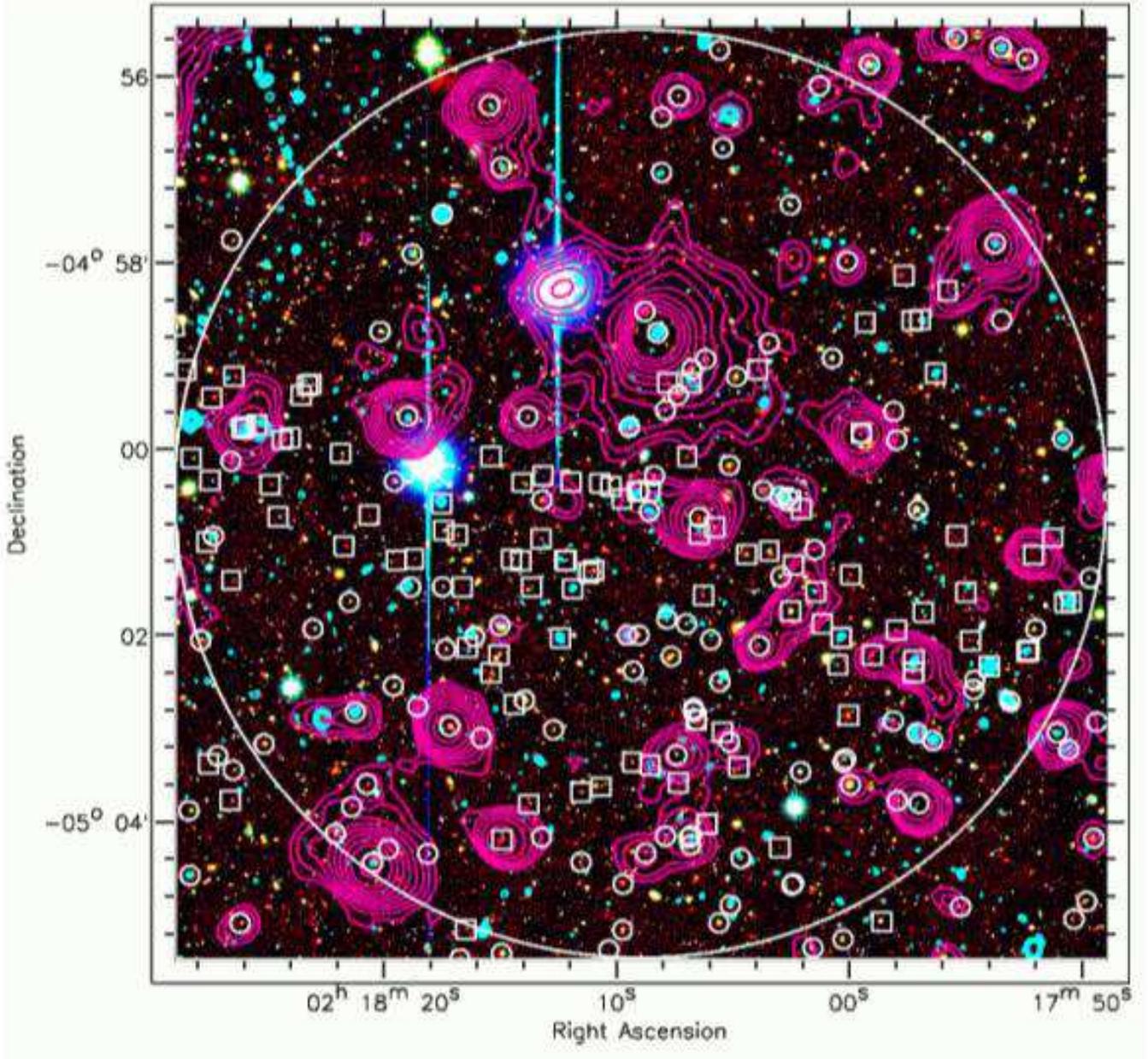}
\caption{\small $Bi^{\prime} K$ image of cluster field CVB13. Symbols
  and contours are as in Fig.~\ref{overlay6}.}
\label{overlay13}
\end{figure*}

Cluster field CVB13 has been described in detail in VB07, where we discussed
the two overdensities found in the DEIMOS data at $z=1.28$ and
$z=1.45$. In this paper, we focus on the cluster at $z=1.28$ as the
data on the second structure are sparse. We will refer to the cluster
at $z=1.28$ as CVB13A. The table of cluster galaxies can be found in
VB07, table 1: galaxies CVB13\_2 to CVB13\_11 are part of CVB13A.

\begin{figure}
\hspace{-1cm}
\includegraphics[width=95mm]{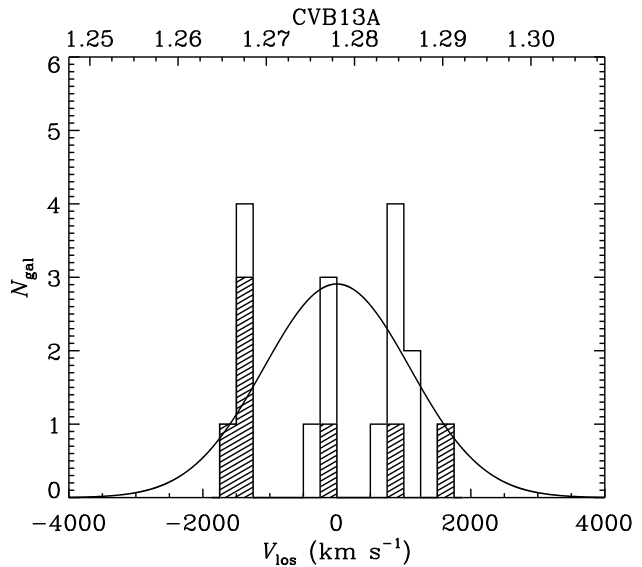}
\vspace{-0.5cm}
\caption{\small Velocity distributions of cluster CVB13A. The
  shaded bins are objects from the SXDF master list. The
  overplotted Gaussian function is determined by $\overline z =
  1.278,\pm\,0.002$, $\sigma_v = 1092\,\pm\,141\,\rm km\,s^{-1}$. The
  data obviously deviate significantly from the Gaussian
  approximation.}
\label{cvb13_veldisp}
\end{figure}

We note that the combination of the DEIMOS and GMOS data with the
SXDF spectroscopic master list yields a slightly different
velocity distribution for CVB13A than the one presented in
VB07, as is shown in Fig.~\ref{cvb13_veldisp}. The mean
redshift and velocity dispersion of the complete sample are $\overline z =
1.278,\pm\,0.002$, $\sigma_v = 1092\,\pm\,141\,\rm km\,s^{-1}$. Like
CVB11B, the velocity distribution appears broad and non-Gaussian, and
therefore the cluster is unlikely to be virialised. It is probable
that the cluster comprises several merging sub-clumps; however the
data do not support a good double Gaussian fit. We perform the same
Bayesian likelihood test on the velocity distribution as on CVB11B
(see Section 3.2.1); however owing to the small sample size we obtain
the same inconclusive result. 

\section{The cluster AGN population}\label{section_agn}
\subsection{X-ray and radio sources}

\begin{figure*}
\begin{center}
\includegraphics[width=150mm,height=95mm]{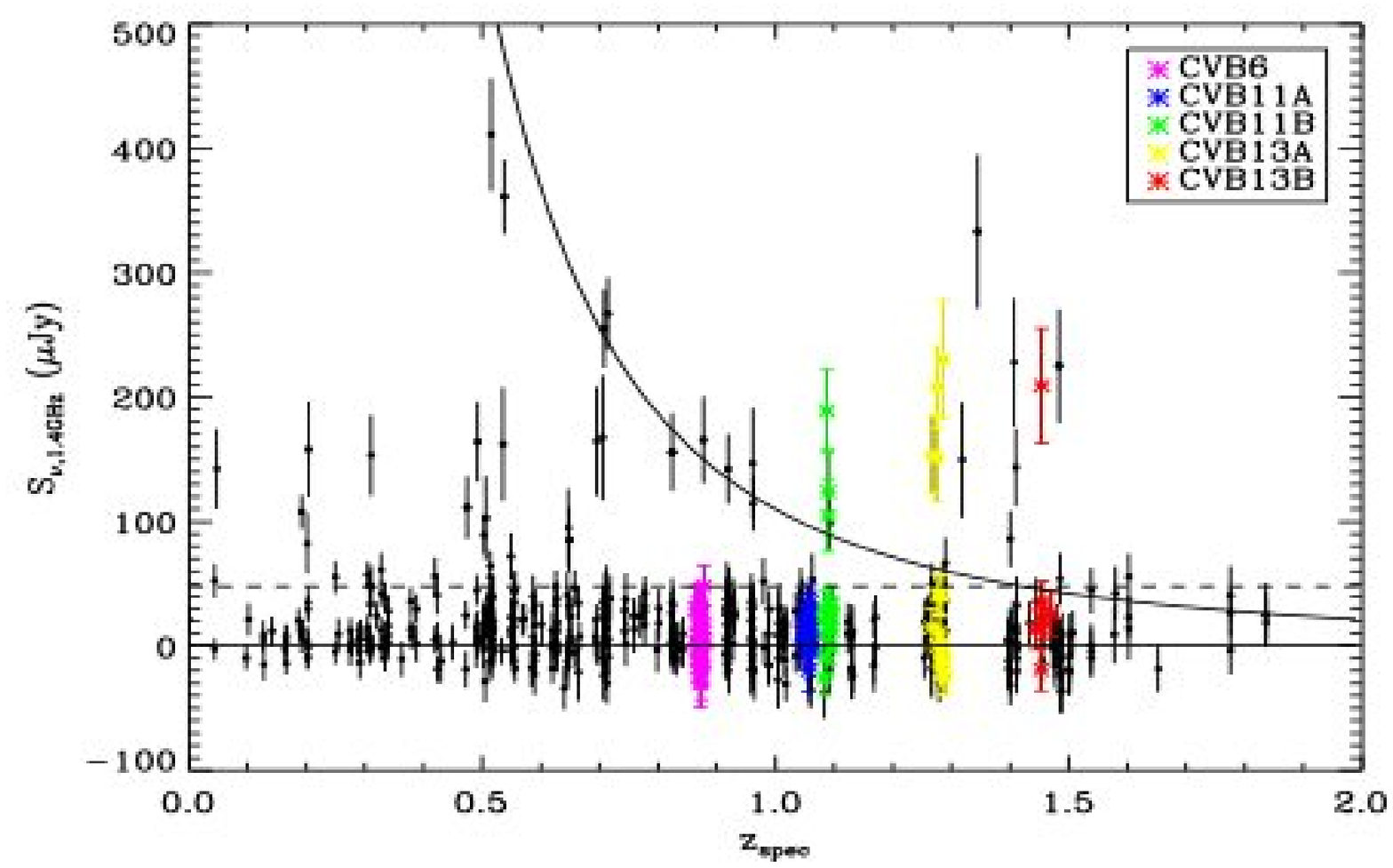}
\end{center}
\vspace{-0.5cm}
\caption{\small Radio flux density at 1.4 GHz versus redshift for all
  objects with a spectroscopic redshift in cluster fields CVB6, CVB11,
  and CVB13. Field galaxies are plotted in black, and cluster galaxies
  are plotted in blue for CVB6, green for CVB11A, yellow for CVB11B,
  and red for CVB13A. Note that two radio sources that are cluster
  members of CVB11B are not visible in this plot; their flux densities
  are $800$ and $2800\,\mu \rm Jy$. The solid curve shows the
  predicted radio flux density caused by a star-burst of $500 \,\rm
  M_{\odot}\,yr^{-1}$ (total over all stellar masses); objects above
  the curve are assumed to be AGN.  The dashed line is the average
  $5\sigma$ flux limit of the radio catalogue.\bigskip}
\label{radio_flux}
\end{figure*}

The deep X-ray and radio data in the SXDF combined with the optical
spectroscopy allow us to investigate the AGN activity in the three cluster
fields, and in particular in the four clusters themselves.

 The X-ray catalogue contains 48 X-ray point sources in our fields, of
 which 42 (88 per cent) have an associated redshift in our
 spectroscopic catalogue. Visual inspections of X-ray-contour overlays
 show that despite $\sim 1.5\as$ scale positional errors on the X-ray
 positions, there is almost always only one plausible candidate for
 follow-up spectroscopy. However in a few cases unambiguous
 identification is impossible. 

The X-ray population at fluxes of $F_{\rm X} > 10^{-18.5}\,\rm
W\,m^{-2}$ in the 0.1 - 10 keV band consists predominantly of AGN,
both obscured (up to $F_{\rm X} \sim 10^{-17} \,\rm W\,m^{-2}$) and
unobscured (e.g. Barger et al. 2001, 2003; Szokoly et al. 2004). At
lower fluxes, a population of star-forming galaxies emerges
(Hornschemeier et al. 2000; Rosati et al. 2002; Norman et
al. 2004). Other sources of X-ray emission from galaxies are X-ray
binaries and the hot interstellar medium (ISM, e.g. Sivakoff, Sarazin \& Irwin
2003). To determine whether the X-ray objects found in our fields
could be star-forming galaxies, we calculate their rest-frame X-ray
fluxes in the $0.1-10$ keV band. However, as the X-ray catalogue gives
total fluxes in the $0.2-12$ keV band, we have to take both the
redshifting of the spectrum ($k$-correction) and the difference in
bands into account to obtain the correct fluxes. Guided by Ueda et
al. (2003), we determine the corrections by assuming an X-ray SED of
the form: \be {\rm SED}\,{\rm [keV]} = E^{-\Gamma} e^{\frac{E}{E_c}},
\ee where $\Gamma = 1.9$, and $E_c = 500 \rm \, keV$. The corrected
flux in the $0.1-10$ keV band then becomes: \be F_{\rm rest,
  0.1-10\,keV} = \frac{\int_{0.1}^{10} {\rm
    SED}\,dE}{\int_{(z+1)0.2}^{(z+1)12}{\rm SED}\,dE} F_{\rm obs,
  2-12\,keV}
\label{fluxcor}
\ee The minimum flux in this band observed in our fields is $F_{\rm
  0.1-10\,keV} = 4.7 \times 10^{-18}\,\rm W\,m^{-2}$, which implies
the X-ray sources in our sample are not star-forming galaxies. To rule
out contaminants by X-ray binaryies and the ISM, we follow the method
of Sivakoff et al. (2008). We calculate the broadband X-ray luminosity
in the 0.3 - 8.0 keV band, and compare with the $K_S$-band luminosity
of the galaxies. We find our sources have X-ray luminosities in the
range of $3 \times 10^{42} < L_{\rm X,0.3-8.0} < 4 \times 10^{46}~\rm
erg~s^{-1}$. The $K_S$-band luminosities are derived from the $K$-band
luminosites from the Ultra Deep Survey (UDS), and band corrected by
subtracting a value of 0.017 (Hewett et al. 2006); they range between
$1 \times 10^{10}$ and $2 \times 10^{12}~\rm L_{K_S,\odot}$. It
appears the large X-ray luminosity of all our sources compared to
their $K$-band luminosity rules out contamination by X-ray binaries
and the ISM. We therefore can safely assume all of the
objects in our sample are AGN.

 The VLA A-array catalogue includes all sources with a flux greater
 than $5\sigma$, where $\sigma$ is the local noise; on average this
 means the sources have $S_{\rm 1.4\, GHz} \grtsim 50 \,\rm\mu
 Jy$. This catalogue contains 87 sources within the three cluster
 fields, of which 40 have a spectroscopic redshift (46 per
 cent). Extragalactic radio sources fall into two main types of
 objects: star-forming galaxies and AGN. Generally the radio emission
 of the brightest sources is caused by an active nucleus, whereas the
 star-forming galaxies dominate the radio population at lower radio
 power. To distinguish between these two populations, we assume that
 the maximum total star-formation rate of a galaxy is $500 \rm\,
 M_{\odot}\,yr^{-1}$ (at which Mauch \& Sadler [2007] find that the
     space density of AGN is $\sim$ 20 higher than star-forming
     galaxies) and calculate the corresponding radio flux density at
     redshifts $0 < z < 2$ using the following relations from Condon
     (1992):

\be 
{\rm SFR}_{\rm non-thermal} = \frac{ P~ {\rm[ W\,Hz^{-1}]}}{5.3\times 10^{21}
 \, \nu^{-\alpha}},
\label{sfr_nonthermal}
\ee

\be
{\rm SFR}_{\rm thermal} = \frac{ P~ {[\rm W\,Hz^{-1}]}}{5.5\times 10^{20}
  \,\nu^{-0.1}}.
\label{sfr_thermal}
\ee

\begin{figure*}
\hspace{-1.5cm}
\includegraphics{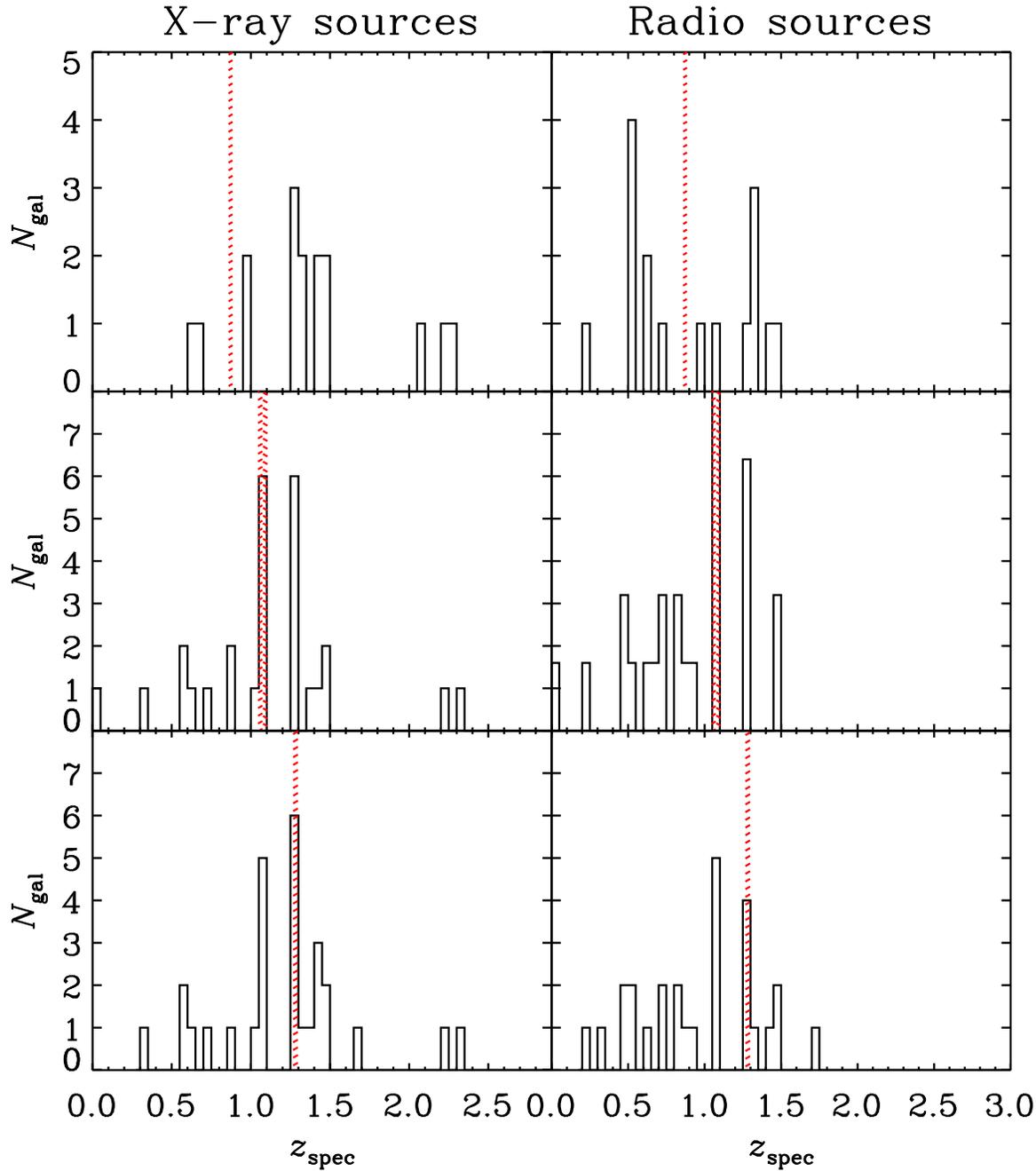}
\caption[Redshift distribution of the radio sources and X-ray point
  sources in the cluster fields of CVB6, CVB11, and CVB13]{\small
  Redshift distribution of the radio sources and X-ray point sources
  in the cluster fields of CVB6 ({\it top}), CVB11 ({\it middle}), and
  CVB13 ({\it bottom}). The red shaded regions are within a velocity
  range of $\pm\,2000\,\rm km\,s^{-1}$ of the clusters.\bigskip}
\label{xrayradio_zhist}
\end{figure*}

Here $P$ is the radio power at frequency $\nu$ (1.4 GHz), and $\alpha
\sim 0.8$ is the non-thermal spectral index. These equations determine
the radio power caused by a star formation in stars of masses $\geq
5\,\rm M_{\odot}$ only; assuming a Salpeter IMF the star-formation
rate in all stars is a factor of 5 higher. This means that the radio
power limit for star-forming galaxies is $P_{\rm lim,\, 1.4\, GHz} =
5 \times 10^{23}\,\rm W\,Hz^{-1}$. Fig.~\ref{radio_flux} shows the limiting
radio flux density versus redshift with the flux densities of all our
radio sources overplotted. These flux densities have been
$k$-corrected to reflect the rest-frame 1.4 GHz radio power; for this
we assume a spectral index of 0.8. All objects with a radio flux
density greater than the limiting radio flux density are assumed to be
AGN; from Fig.~\ref{radio_flux} we can see that all objects we find with a
flux density $\geq 5\sigma$ at our clusters' redshifts fall within
this category.

\subsection{The number density of active galaxies}
The redshift distribution of the X-ray and radio sources is shown in
Fig.~\ref{xrayradio_zhist} for each of the fields; note that a number
of sources is included in more than one histogram due to the overlap
of the cluster fields (see Fig.\ref{fields}).

\begin{figure*}
\begin{center}
\includegraphics[width=150mm]{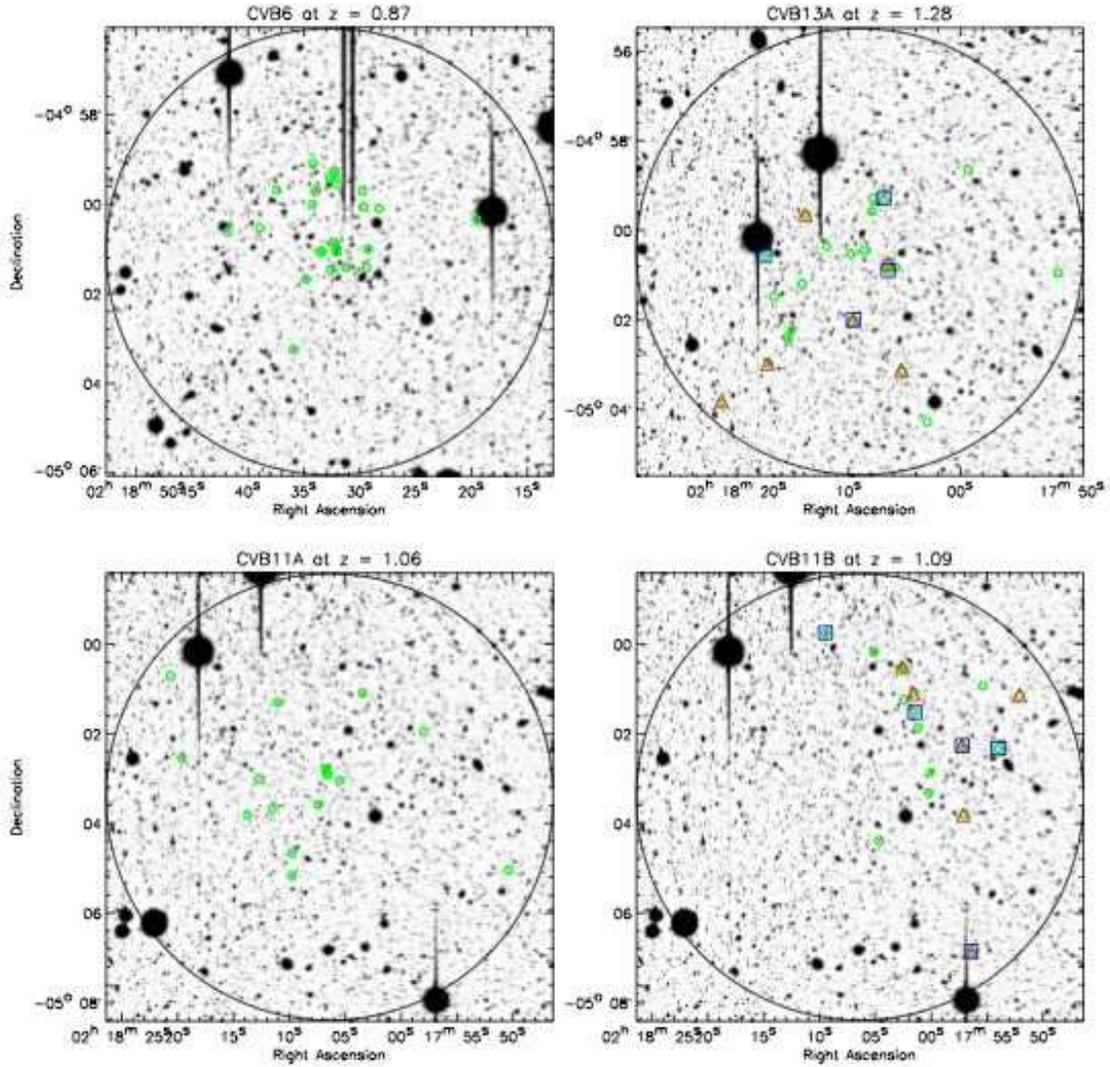}
\end{center}
\vspace{0.5cm}
\caption{\small Cluster galaxies (green circles) and associated X-ray
  (red triangles) and radio (blue squares) AGN overlaid on
  $i^{\prime}$-band images of CVB6, CVB11A, CVB11B, and CVB13A. All
  objects marked by symbols are within $\pm\,2000\,\rm km\,s^{-1}$ of
  the cluster redshift. The black circle marks the $5\am$-radius field
  that was investigated. Note that there are no AGN associated with
  clusters CVB6 and CVB11A.\bigskip}
\label{agn}
\end{figure*}

It is apparent from this figure that there are no radio
sources nor X-ray point sources associated with cluster CVB6. For
clusters CVB11A, CVB11B, and CVB13A we inspect the velocity range of
$\pm\,2000\,\rm km\,s^{-1}$ around the cluster redshifts: any radio or
X-ray point sources in the respective fields within this velocity
interval are taken to be associated with the clusters. Cluster CVB11A
also has no associated radio or X-ray point sources; CVB11B and CVB13A
however both contain a number of X-ray and radio sources. The
positions of the cluster galaxies, X-ray point sources, and radio
sources are plotted for each cluster in
Fig.~\ref{agn}. Table~\ref{table_agn} lists the number of associated
AGN per cluster. We note that the broadband X-ray luminosity limit for CVB6 and
CVB11A is $2 \times 10^{42}~\rm erg~s^{-1}$; this means we may be
missing low-luminosity X-ray AGN of $L_{\rm X} \ge 10^{41}~\rm
erg~s^{-1}$. However, the luminosity limits of CVB11B and CVB13 are
even higher as they lie at greater redshift. Therefore, we would be
missing the same or even larger fraction of AGN in these clusters and
the lack of detected AGN in the lower-redshift clusters is not due to an
observational bias.

\begin{table*}
\vspace{0.2cm}
\begin{center}
\begin{tabular}{lccccccc}
\hline 
\noalign{\smallskip} 
\multicolumn{1}{l}{Cluster}                     & 
\multicolumn{1}{c}{$N_{\rm gal}$}                & 
\multicolumn{1}{c}{$N_{\textrm X-ray}$}              &
\multicolumn{1}{c}{$N_{\textrm radio}$}              &
\multicolumn{1}{c}{$N_{\textrm exp,\, X-ray}$}        &
\multicolumn{1}{c}{$N_{\textrm exp,\, radio}$}        &
\multicolumn{1}{c}{$P(N_{\textrm X-ray}|N_{\textrm exp})$}        &
\multicolumn{1}{c}{$P(N_{\textrm radio}|N_{\textrm exp})$}        \\
\noalign{\smallskip} 
\hline 
\noalign{\smallskip} 
CVB6   & 25 & 0 & 0 & 0.92 & 0.15 & $3.5 \times 10^{-1}$ & $7.0 \times 10^{-1}$ \\
CVB11A & 14 & 0 & 0 & 0.90 & 0.16 & $3.6 \times 10^{-1}$ & $6.9 \times 10^{-1}$ \\
CVB11B & 16 & 6 & 5 & 0.90 & 0.16 & $2.7 \times 10^{-4}$ & $1.0 \times 10^{-5}$ \\
CVB13A & 18 & 5 & 4 & 0.85 & 0.15 & $1.7 \times 10^{-3}$ & $1.0 \times 10^{-5}$ \\
\noalign{\smallskip}
\hline 
\end{tabular}  
\end{center}
\caption{\small The numbers of AGN associated with the clusters in a
  field of $5\am$ radius, together with the expected numbers. The
  first column is the cluster ID, the second is the total number of
  spectroscopically confirmed cluster galaxies, and columns 3 and 4
  are the number of X-ray and radio AGN respectively. Columns 5 and 6
  show the numbers of expected X-ray and radio AGN if the inspected
  cluster fields were random background fields. Columns 7 and 8 are
  the probabilities that the number of observed AGN are caused by the
  background distribution of AGN. \bigskip}
\label{table_agn}
\end{table*}

We can calculate the number of expected X-ray and radio AGN within the
probed volume, by integrating over the respective luminosity
functions. For the X-ray sources, we use the Hard X-ray Luminosity
Function (HXLF) in the 2-10 keV band from Ueda et al. (2003). This is
a luminosity-dependent density evolution model of the following form:
\be d\Phi = A\Bigl\{\Bigl(\frac{L_{\rm X}}{L^*}\Bigr)^{\gamma_1} +
\Bigl(\frac{L_{\rm X}}{L^*}\Bigr)^{\gamma_2}\Bigr\}^{-1}
e(z)d\log(L_{\rm X}),
\label{xrayphi}
\ee
where $\Phi$ is the number density per cubic Mpc, and $e(z)$ is the
evolution factor. The values of the constant parameters in this and
the following two equations are given in Ueda et al. (2003). The
evolution factor is given by:
\be
e(z) = \left\{
\begin{array}{lr}
(1+z)^{p_1} & z < z_c (L_{\rm X})\\
e(z_c)\Bigl( \frac{1+z}{1+z_c(L_{\rm X})}\Bigr)^{p_2} & z \geq
z_c(L_{\rm X})\\
\end{array}
\right.
\ee
Here $z_c$ is the cut-off redshift above which the evolution
terminates, which is dependent on the X-ray luminosity in the
following way:
\be
z_c(L_{\rm X}) =  \left\{
\begin{array}{lr}
z_c^* & L_{\rm X} \geq L_{\rm a}\\
z_c^* \Bigl(\frac{L_{\rm X}}{L_{\rm a}}\Bigr)^{\alpha} & L_{\rm X} < L_{\rm a}\\
\end{array}
\right .
\label{zc}
\ee

We calculate the minimum observed flux in the $2-10$ keV band,
applying the corrections of Eq.~\ref{fluxcor}, and obtain $F_{\rm
  min,\,2-10\,keV} = 4.1 \times 10^{-19}\,\rm W~m^{-2}$. Converting
this to a luminosity, integrating Eq.~\ref{xrayphi} from this value
upwards, and multiplying by the volume given by the circular field of
$5\am$-radius and $\Delta V = 4000\,\rm km\,s^{-1}$ gives on average
one expected X-ray source in a random volume of this size. Taking the
completeness into account, this number is reduced to 0.9.

The radio luminosity function consists of three components: (i) a
luminosity function for radio-loud AGN (Willott 2001); (ii) a radio
luminosity function for radio-quiet AGN derived from the X-ray
luminosity function from Ueda et al. (2003) and converted to radio
using the relations set out in Brinkmann et al. (2000); (iii) a
luminosity function for star-forming galaxies derived from infrared
observations (Yun, Reddy \& Condon 2001), taking into account the
redshift evolution observed in submillimetre source counts (Blain
1999). Combining all three, integrating from our luminosity limit
upwards, and accounting for our completeness, yields an average
expected number of 0.2 radio sources in our cluster fields if they
were random fields (see Jarvis \& Rawlings 2004 and Wilman et al. 2008
for a detailed description of the method used). The exact expected
number of AGN per cluster are given in Table~\ref{table_agn}. We use
Poissonian low-number statistics to calculate the probability that the
observed numbers of AGN are fluctuations of the expected background
model. These numbers are also listed in Table~\ref{table_agn}; we
conclude that the observed numbers of X-ray and radio sources in
CVB11B and CVB13A are $3-5\sigma$ away from the expected numbers,
which indicates that the AGN in these fields are clustered to a
highly-significant level. The absence of AGN in CVB6 and CVB11A is
consistent with the AGN population being no different in these
clusters than in the field.

\section{Discussion}\label{conc6}
To inspect the two-dimensional clustering of the AGN associated with
CVB11B and CVB13A, we calculate the $A/A_{\rm max}$ statistic for both
the X-ray and the radio sources. This statistic is the ratio of the
average area, in which the AGN occur, to the maximum investigated
area. For each AGN the area is defined as the circle with a radius
determined by the distance of the AGN to the cluster centre. $A_{\rm
  max}$ is the area of the circle with a radius of 5$\am$. If the AGN
are randomly distributed over the field, the value of $A/A_{\rm max}$
is 0.5; a value $< 0.5$ indicates clustering in right ascension and
declination. The result is shown in Table~\ref{table_amax}; evidently,
the AGN are clustered within a smaller field than the total
$5\am$-radius fields. This is not surprising, as the $r_{200}$ virial
radius of a cluster such as CVB6 ($M_{200} = 1.6 \times 10^{14} \,\rm
M_{\odot}$) is only 1.3 Mpc in proper coordinates, whereas a field of
$5\am$ radius would correspond to 2.4 Mpc at $z=1.0$.
\begin{table}
\bigskip
\begin{center}
\begin{tabular}{lccc}
\hline 
\noalign{\smallskip} 
\multicolumn{1}{l}{Cluster}                    & 
\multicolumn{1}{c}{$A_{\textrm X-ray} / A_{\textrm max}$} &  
\multicolumn{1}{c}{$A_{\textrm radio} / A_{\textrm max}$} &
\multicolumn{1}{c}{$A_{\textrm AGN} / A_{\textrm max}$} \\
\noalign{\smallskip} 
\hline 
\noalign{\smallskip} 
CVB11B &  0.2$\,\pm\,$0.2  & 0.3$\,\pm\,$0.3 & 0.2$\,\pm\,$0.2 \\
CVB13A &  0.3$\,\pm\,$0.3  & 0.1$\,\pm\,$0.3 & 0.2$\,\pm\,$0.2  \\
\noalign{\smallskip}
\hline 
\end{tabular}  
\end{center}
\caption{\small The $A/A_{\rm max}$ statistic for both the X-ray and
  the radio sources associated with clusters CVB11B and
  CVB13A. Columns 2 and 3 show the statistic for X-ray and radio
  sources respectively, and column 4 is the combined statistic after
  removal of coincident detections.}
\label{table_amax}
\end{table}

\begin{table*}
\begin{center}
\begin{tabular}{lccccccc}
\hline
\noalign{\smallskip} 
\multicolumn{1}{l}{Cluster}                     & 
\multicolumn{1}{c}{$N_{\textrm X-ray}$}              &
\multicolumn{1}{c}{$N_{\textrm radio}$}              &
\multicolumn{1}{c}{$N_{\textrm exp,\, X-ray}$}        &
\multicolumn{1}{c}{$N_{\textrm exp,\, radio}$}        &
\multicolumn{1}{c}{$P(N_{\textrm radio}|N_{\textrm exp})$}        &
\multicolumn{1}{c}{$P(N_{\textrm radio}|N_{\textrm exp})$}        \\
\noalign{\smallskip} 
\hline 
\noalign{\smallskip} 
CVB6   & 0 & 0 & 7.6 & 1.3 & $5.1 \times 10^{-4}$ & $2.6 \times 10^{-1}$ \\
CVB11A & 0 & 0 & 5.6 & 0.98 & $3.0 \times 10^{-3}$ & $3.3 \times 10^{-1}$ \\
CVB11B & 5 & 3 & 5.3 & 0.96 & $6.1 \times 10^{-1}$ & $7.2 \times 10^{-2}$ \\
CVB13A & 2 & 4 & 3.9 & 0.70 & $2.5 \times 10^{-1}$ & $5.6 \times 10^{-3}$ \\
\noalign{\smallskip}
\hline 
\end{tabular}  
\end{center}
\caption[The numbers of AGN associated with clusters CVB6, CVB11A+B,
    and CVB13A within a radius of $r_{200}$, together with the
    expected numbers]{\small The numbers of AGN associated with the
    clusters within a radius of $r_{200}$ of CVB6 ($2.6\am$), together
    with the expected numbers of AGN assuming a cluster with an
    overdensity of factor 200. The first column is the cluster ID, and
    columns 2 and 3 are the number of X-ray and radio AGN
    respectively. Columns 4 and 5 show the numbers of expected X-ray
    and radio AGN, assuming their clustering traces the mass
    overdensity. Columns 6 and 7 are the probabilities that the number
    of observed AGN are caused by an overdensity of factor 200.}
\label{table_agn2}
\vspace{-0.5cm}
\end{table*}

As CVB11B and CVB13A do not appear to be virialised, we cannot
calculate a virial mass -- and thus radius -- from the velocity
dispersion. The number of galaxies found in the clusters suggest
however that they are of lower mass than CVB6 and therefore confined
to a smaller radius. On the other hand, if the clusters are not
virialised yet, they could occupy a larger volume than virialised
systems of the same mass. We therefore assume these effects cancel out
roughly, and examine the AGN of CVB11B and CVB13A within the virial
radius $r_{200} = 1.3 \rm \, Mpc$ (proper coordinates) of CVB6, which
corresponds to $2.6\am$ at the redshift of the two clusters. We
determine the number of AGN in these new fields and show them in
Table~\ref{table_agn2}. Further, we recalculate the number of expected
X-ray and radio sources, this time assuming a cluster environment (in
proper coordinates) with a total overdensity of a factor of 200
(implied by the definition of $r_{200}$). Using these new numbers, we
can establish the probability that the number of observed AGN is
caused by the overdensity in the background distribution. These
numbers are also given in Table~\ref{table_agn2}. An interesting
result emerges: the lack of X-ray sources in CVB6 and CVB11A is
significant at a level of $\geq 3\sigma$, whereas the lack of radio
sources is not significant. Contrarily, the number of X-ray sources in
CVB11B and CVB13A are consistent with being caused by an overdensity
of factor 200, whereas the radio sources appear to be even more
clustered than that. In fact, the numbers suggest the radio sources
are a factor 3 - 6 more clustered than the X-ray sources.

In summary, we have presented evidence that the AGN population of
clusters appears to change fundamentally during the evolution of the
cluster, although our conclusions are limited by small number
statistics. Clusters CVB11B and CVB13A seem to be in a state of
pre-virialisation, as can be derived from their velocity
distributions. They show a number of associated AGN far above the
background level, and consistent with an overdensity comparable with
the total mass overdensity, although the radio galaxies appear to be
even more heavily clustered. Clusters CVB6 and CVB11A are in a later
evolutionary stage, and both have an extended X-ray detection. CVB6 is
the best example of this: the X-ray properties and velocity
distribution all lie neatly on normal cluster relations. These two
clusters have few, if any, associated AGN, which means that the AGN
activity is less or equal to that of the galaxy field. It is possible
that as the cluster virialises, AGN activity is extinguished, leaving
the clusters quiescent.

A potential explanation for this observed phenomenon is as
follows. X-ray AGN contain a supermassive black hole in their galactic
nucleus which accretes gas at a high rate. This means a large amount
of fuel is needed on small scales, which could be caused by a
galaxy-galaxy merger (e.g. Barnes \& Hernquist 1996). Pre-virialised
clusters most likely consist of merging sub-groups with low internal
velocity dispersions, which allows galaxy mergers. The galaxies in
virialised clusters however have high relative velocities, which
suppresses the galaxy merger rate (e.g. Giovanelli \& Haynes
1985). Hence the X-ray AGN fraction is much lower in virialised
clusters than in systems which are in an earlier evolutionary stage.

Radio AGN are probably caused by rapidly spinning supermassive black
holes in the nuclei of massive galaxies, created my the merger of two
nuclear black holes of similar mass (Wilson \& Colbert
1995). Objects at a given radio luminosity can have a wide range of
accretion rate of the black hole. The high-accretion sources are
identical to the X-ray sources; indeed we observe some overlap between
our X-ray and radio AGN. In a cluster environment galaxy-galaxy
harassment can boost the accretion rate (Moore et al. 1996), which
increases the radio power (e.g. Willottt et al. 1999). Also, the
luminosity of a radio jet can be increased by higher intergalactic gas
densities, as the jet encounters a denser medium (e.g. Prestage \&
Peacock, 1988; Daly, 1995). The intracluster medium both in virialised
and pre-virialised clusters is generally denser than in a field
environment; the galaxy interaction is higher in pre-virialised
systems for the same reason as for the galaxy merger rate.

Hopkins et al. (2005) show that the lifetime of a luminous quasar
caused by a merger is expected to be of the order of $\sim 10^7\,$yr
($B$-band luminosity greater than $10^{11}\,\rm L_{\odot}$) to
$10^9\,$yr ($B$-band luminosity greater than $10^{9}\,\rm L_{\odot}$)
when taking into account attenuation by obscuring material, with an
intrinsic lifetime of $\sim 10^8\,-10^9$yr.
This means that if no new AGN are triggered after virialisation, the
cluster would be left quiescent after this length of
time. Furthermore, during virialisation it is likely that one big
radio AGN is triggered, that could shut down all AGN activity
henceforward in a cluster (Rawlings \& Jarvis, 2004). This is a
further explanation for the lack of activity in virialised clusters
such as CVB6 and CVB11A, where this event may already have happened,
whereas CVB11B and CVB13A have not encountered this phenomenon yet.

\begin{figure}
\hspace{-1cm}
\includegraphics[width=100mm]{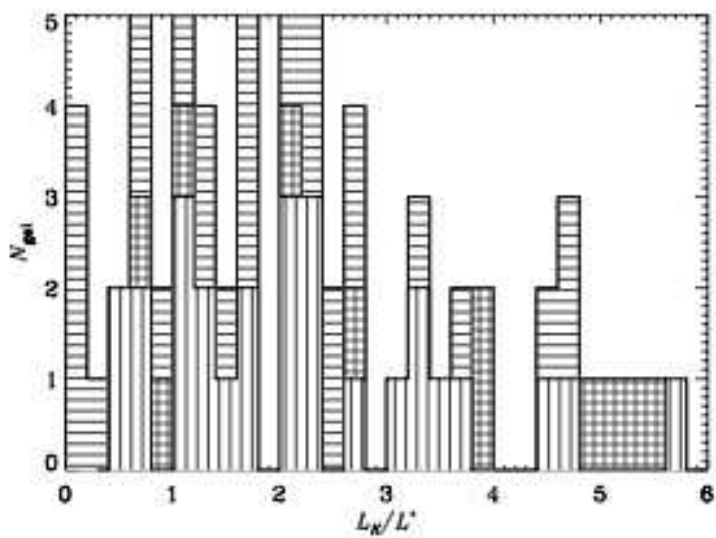}
\caption[The luminosity in terms of $L^*$ for radio and X-ray AGN]{The
  luminosity in terms of the passively evolving $L^*$ for radio
  (vertical hatching) and X-ray (horizontal hatching) AGN. The
  cross-hatched bins are AGN that show both X-ray and radio
  emission.\bigskip}
\label{lfraction}
\vspace{-1cm}
\end{figure}

If the above scenario is valid, we expect the galaxy hosts of the AGN
to be more massive than normal galaxies. The prediction for X-ray AGN
is $\sim 2L^*$, whereas it is slightly higher for radio galaxies ($2-3
L^*$) as the spinning black holes appear to be found only in the
most-massive objects (e.g. Dunlop et al. 2003). Fig.~\ref{lfraction}
shows the $K$-band luminosity histogram of all X-ray and radio AGN in
our fields (at all redshifts) expressed as a fraction of $L^*$
(assumed to be passively evolving, with $M_K^*=-24.18$ at $z=0$ [Cole
  et al. 2001, corrected for cosmology and difference in $K$-bands]),
which is the luminosity at which the break in the galaxy luminosity
function occurs. It appears that the radio galaxies (vertically
hatched bins) are slightly more massive than the X-ray galaxies
(horizontally hatched bins). The median luminosity for the X-ray
sources is $L_K = 1.7\,\pm\,0.7\,L^*$, whereas the for radio sources
we find a median of $L_K = 2.3\,\pm\,0.1\,L^*$. Nonetheless we need to
bear in mind that the higher measured AGN luminosities could instead
be due to an observational selection bias as, given the same Eddington
accretion rate, the more powerful AGN reside in more massive galaxies,
which would be more easily detected.

If the observed distribution of luminosities reflects the true AGN
luminosity distribution, this would support the hypothesis described
above for AGN fractions residing in clusters of different evolutionary
stages. However, we are dealing with low-number statistics and a more
comprehensive sample is needed to confirm our findings. This is of
particular importance, as Martini et al. (2007) show that there is
significant variation in the X-ray selected AGN fraction between
clusters at lower redshift. Ruderman \& Ebeling (2005) find an
overdensity of X-ray AGN in 51 massive clusters at $0.3 < z <
0.7$. Their sample shows an excess in the centre of the clusters,
likely to be caused by the central cluster galaxy, followed by a
depletion in the intermediate regions and a secondary excess at a
distance greater than 2.5 Mpc. The latter is attributed to galaxy
merging and interaction during infall into the cluster. At first
glance, this is at odds with our findings, as we do not find an excess
in our virialised clusters. However, our sample differs significantly
from Ruderman \& Ebeling's, as our clusters are less massive and at
much higher redshift. These circumstances could cause the central
cluster galaxy to not yet have been activated if it is triggered in a
later stage of the cluster's evolution. Furthermore, our field of view
of $5\am$ corresponds to 2.4 Mpc at $z=1$, meaning that we do not
probe the outer regions in which Ruderman \& Ebeling find their
secondary excess. Their conclusion that this is caused by merging
galaxies is actually in agreement with our findings for our
non-virialised clusters. It is evident that to link all studies of AGN
in clusters, we will need large cluster samples imaged in both the
radio and X-ray regime at a range of redshifts. Future deep,
wide-field optical/infrared surveys, such as the Visible and Infrared
Survey Telescope for Astronomy (VISTA) Deep Extragalactic Observations
Survey (VIDEO), coupled with X-ray and radio observations, will be
vital to acquire a large sample of clusters and AGN at $z > 1$.

\section{Acknowledgments}
The authors would like to thank Doi, Morokuma, Miyazaki, Saito,
Yamada, Satoshi, Smail, and Croom for sharing their redshifts in the
SXDF master list. CVB acknowledges support from STFC in the form of a
postdoctoral felloship. This paper is partly based on observations
obtained at the Gemini Observatory, which is operated by the
Association of Universities for Research in Astronomy, Inc., under a
cooperative agreement with the NSF on behalf of the Gemini
partnership: the National Science Foundation (United States), the
Science and Technology Facilities Council (United Kingdom), the
National Research Council (Canada), CONICYT (Chile), the Australian
Research Council (Australia), Ministério da Ciência e Tecnologia
(Brazil) and SECYT (Argentina). Some of the data presented herein were
obtained at the W.M. Keck Observatory, which is operated as a
scientific partnership among the California Institute of Technology,
the University of California and the National Aeronautics and Space
Administration. The Observatory was made possible by the generous
financial support of the W.M. Keck Foundation. The authors wish to
recognize and acknowledge the very significant cultural role and
reverence that the summit of Mauna Kea has always had within the
indigenous Hawaiian community.  We are most fortunate to have the
opportunity to conduct observations from this mountain.

\newpage
\begin{appendix}
\newpage
\section{GMOS targets}

\begin{table*}
\include{table_targets}
\caption{\small Targeted cluster candidates. The IDs (column 1) and
  photometric redshifts (column 2) are from VB06. The RA and Dec
  (column 3 and 4) are the coordinates of the telescope pointing and
  column 5 is the position angle of the instrument (east from
  north). Columns 6, 7, and 8 give the number of target galaxies of
  priority 1, 2, and 3 respectively, and columns 9, 10, and 11 show
  the number of galaxies of each priority included in the MOS mask.}
\label{targets}
\vspace{-0.5cm}
\end{table*}

\begin{table*}
\vspace{0.5cm}
\begin{center}
\begin{tabular}{lcccc}
\hline
ID     & $z_{\rm phot}$    & $z_{\rm spec}$ & $N_{\rm gal}$     & $N_{\rm gal}$     \\
       &                   &                & ($\Delta z=0.02$) & ($\Delta z=0.04$) \\
\hline
CVB6   & 0.76\,$\pm$\,0.12 & 0.87           & 7                 & 7                 \\
CVB7   & 0.78\,$\pm$\,0.06 & 0.91           & 4                 & 5                 \\
CVB9   & 0.80\,$\pm$\,0.06 & 0.92           & 6                 & 12                \\
CVB11  & 0.95\,$\pm$\,0.11 & 1.05           & 6                 & 8                 \\
\hline
\end{tabular}
\end{center}
\caption{\small Spectroscopic redshifts of the five clusters targeted
  with GMOS and DEIMOS. The spectroscopic redshift given in
  column 3 is the peak of the redshift distribution in
  Figs.~\ref{zspec_gmos}. For a further discussion on the exact cluster redshifts
  of CVB6 and CVB11, see Section 3. The number of galaxies
  within bins of $\Delta z=0.2$ and $\Delta z=0.4$ of the redshift
  peak is given in columns 4 and 5 respectively.\bigskip}
\label{clust_z}
\vspace{-0.5cm}
\end{table*}

\begin{figure*}
\begin{center}
\includegraphics[width=6cm]{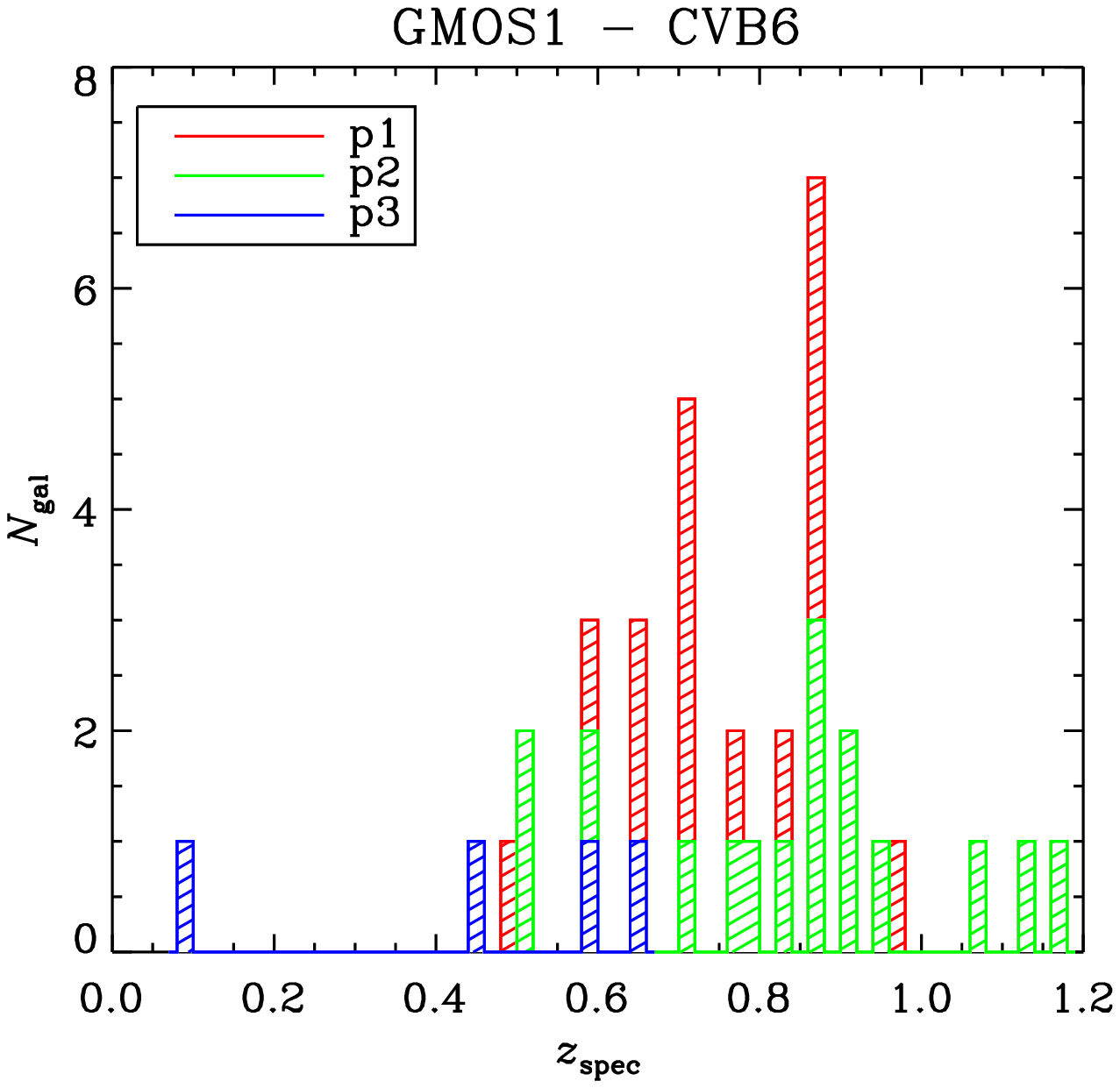}
\hspace{-1cm}
\includegraphics[width=6cm]{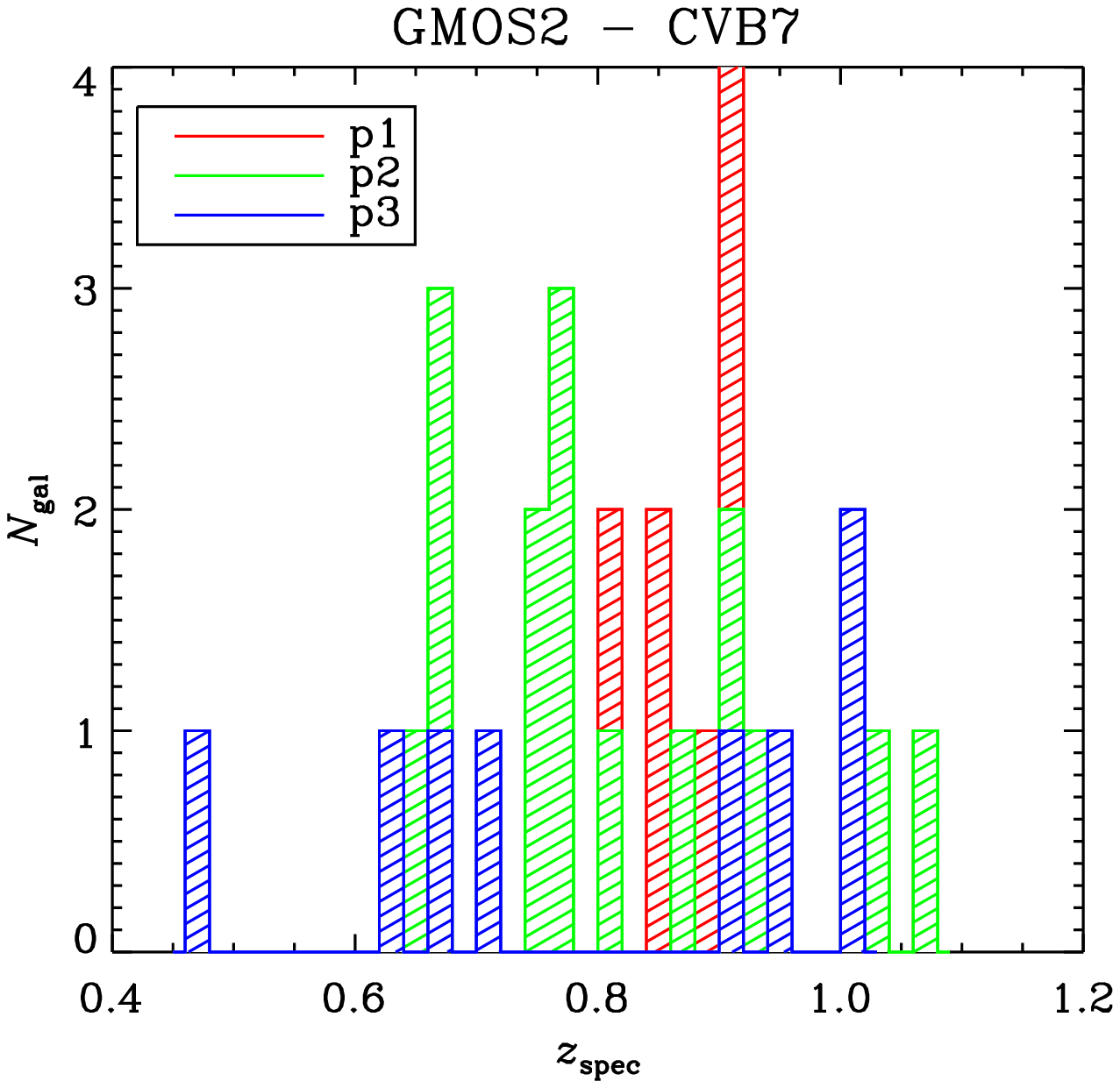}\\
\includegraphics[width=6cm]{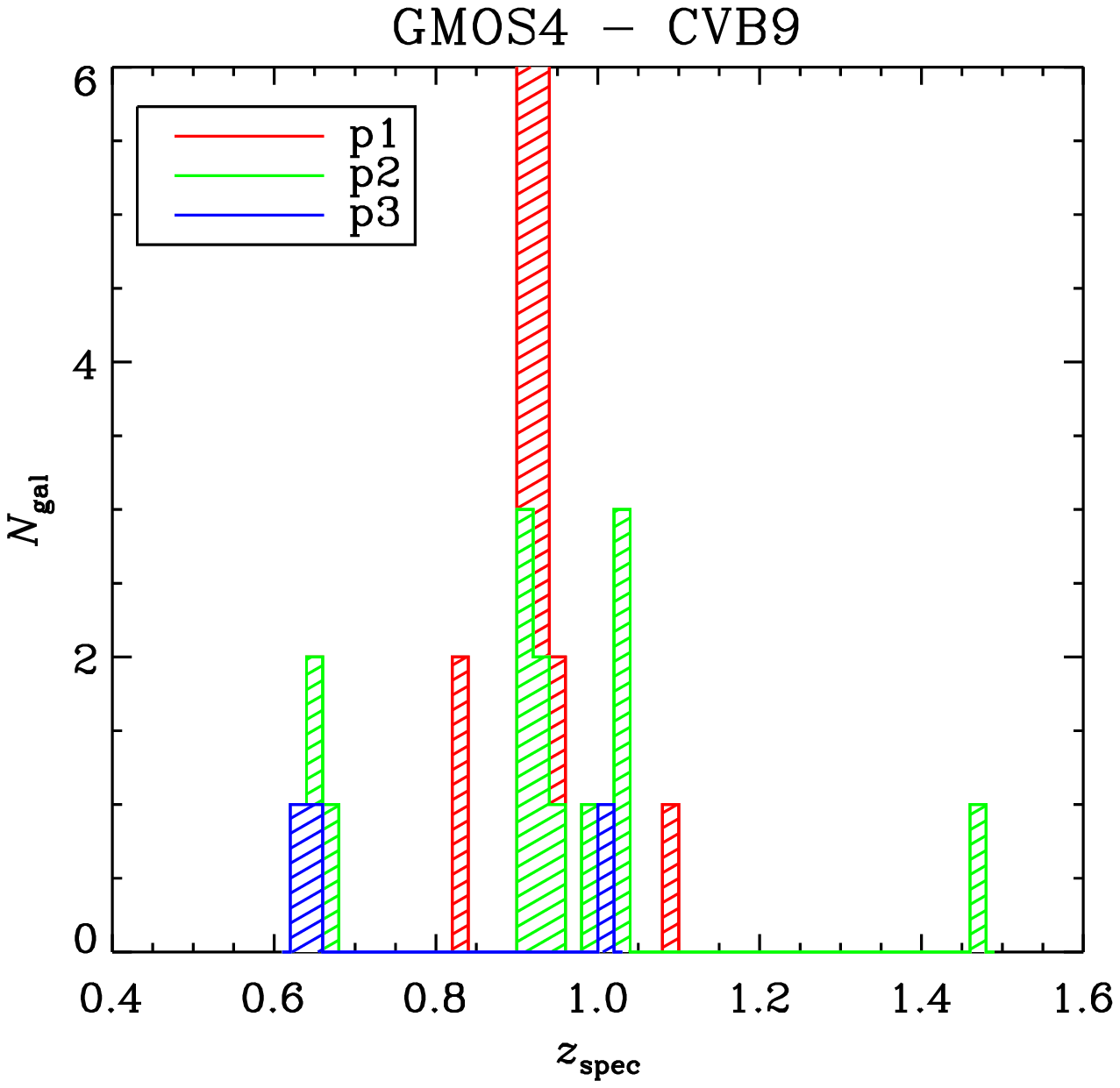}
\hspace{-1cm}
\includegraphics[width=6cm]{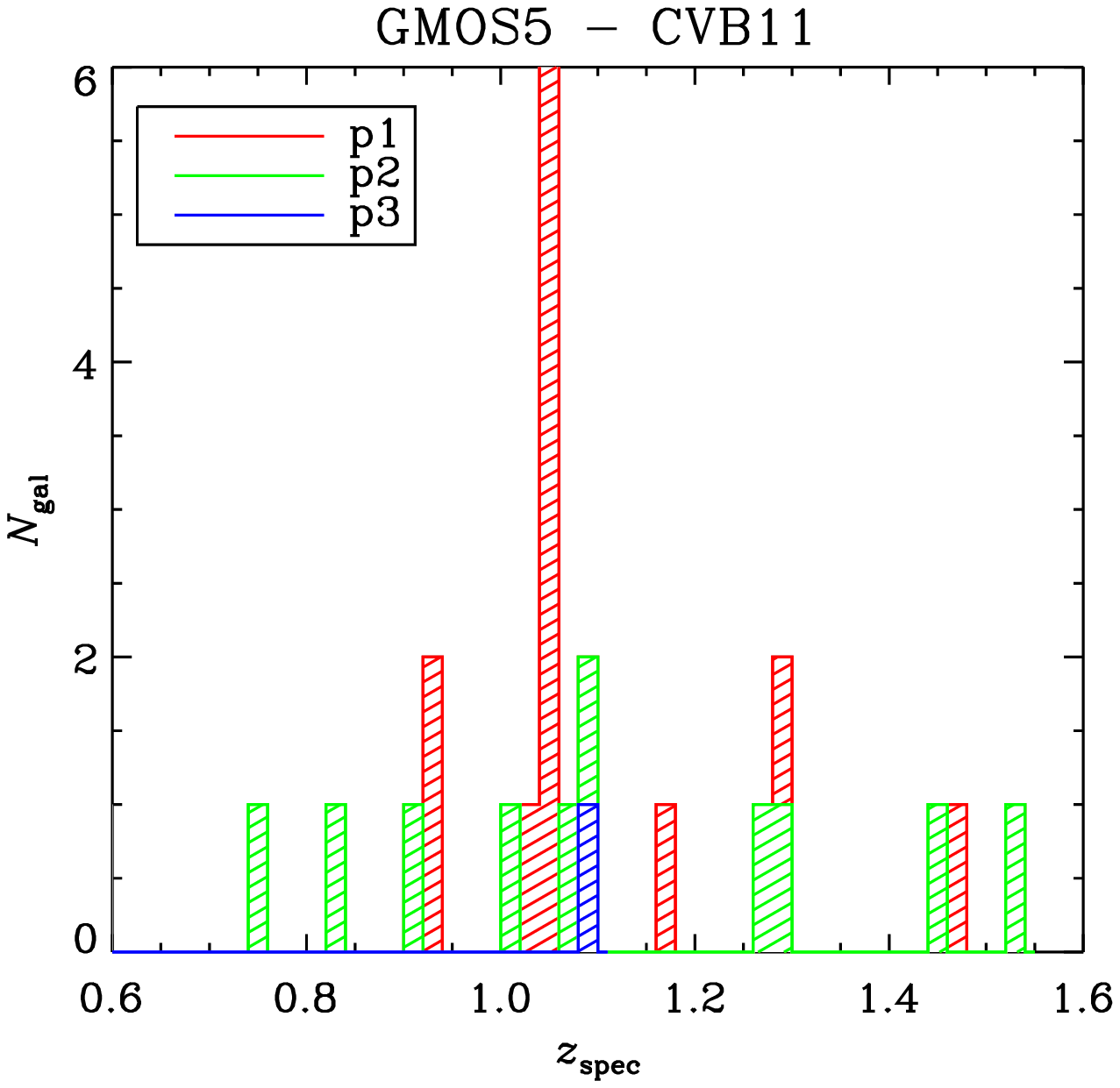}
\end{center}
\vspace{-0.7cm}
\caption{\small Distributions of the spectroscopic redshifts in each of the
  GMOS fields, targeting clusters CVB6, CVB7, CVB9, and CVB11. The
  three initial priorities of the targets (see
  Section 2) are colour-coded: red is priority 1,
  green is priority 2, and blue is priority 3.}
\label{zspec_gmos}
\end{figure*}
\newpage
\section{Photometric versus spectroscopic redshifts}
\begin{figure}
\hspace{-1cm}
\includegraphics[width=90mm]{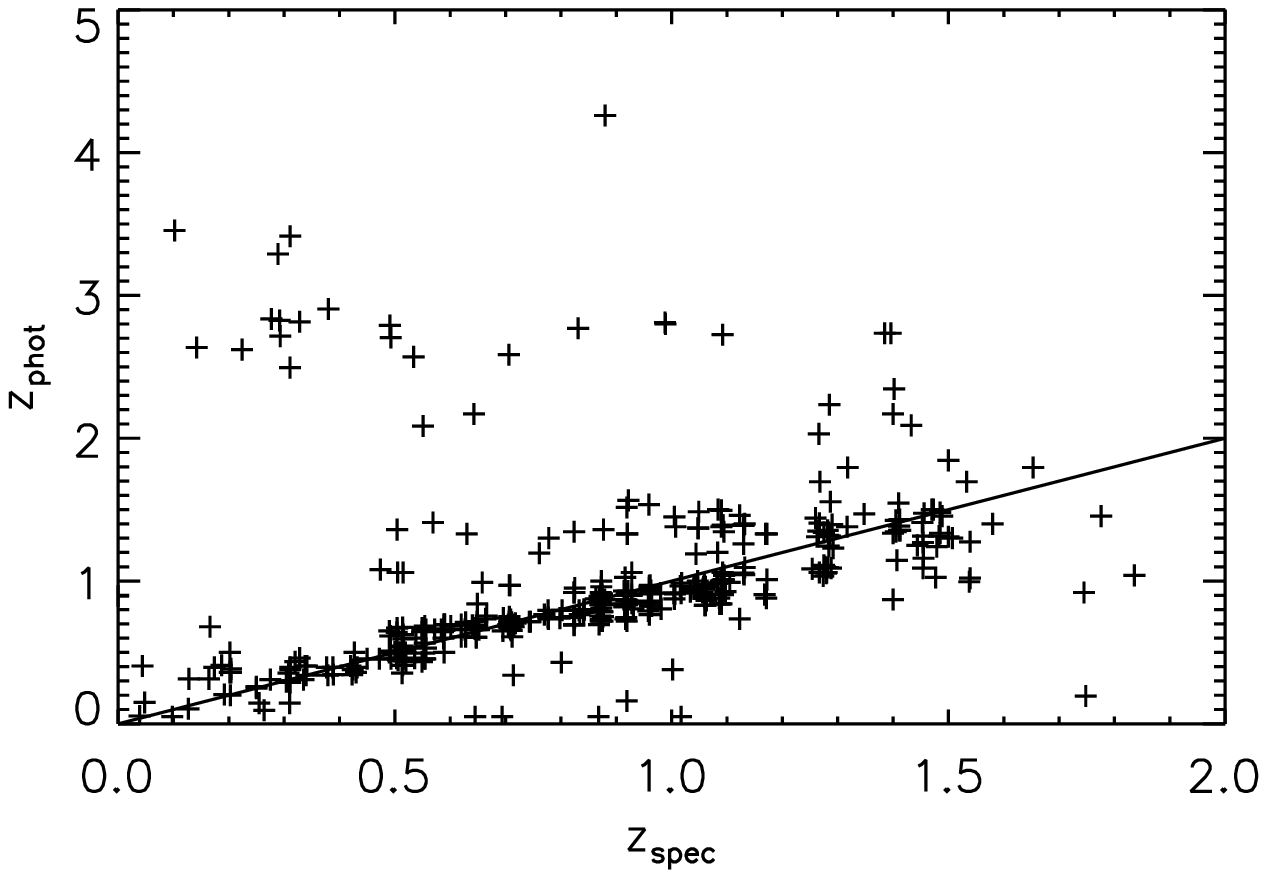}
\caption{\small Photometric redshift versus spectroscopic redshift for
  all galaxies in our sample at $z_{\rm spec} \leq 2$. Overplotted for
  reference is the line for which $z_{\rm phot} = z_{\rm spec}$.}
\label{photovsspec}
\end{figure}

\begin{figure}
\includegraphics[width=90mm]{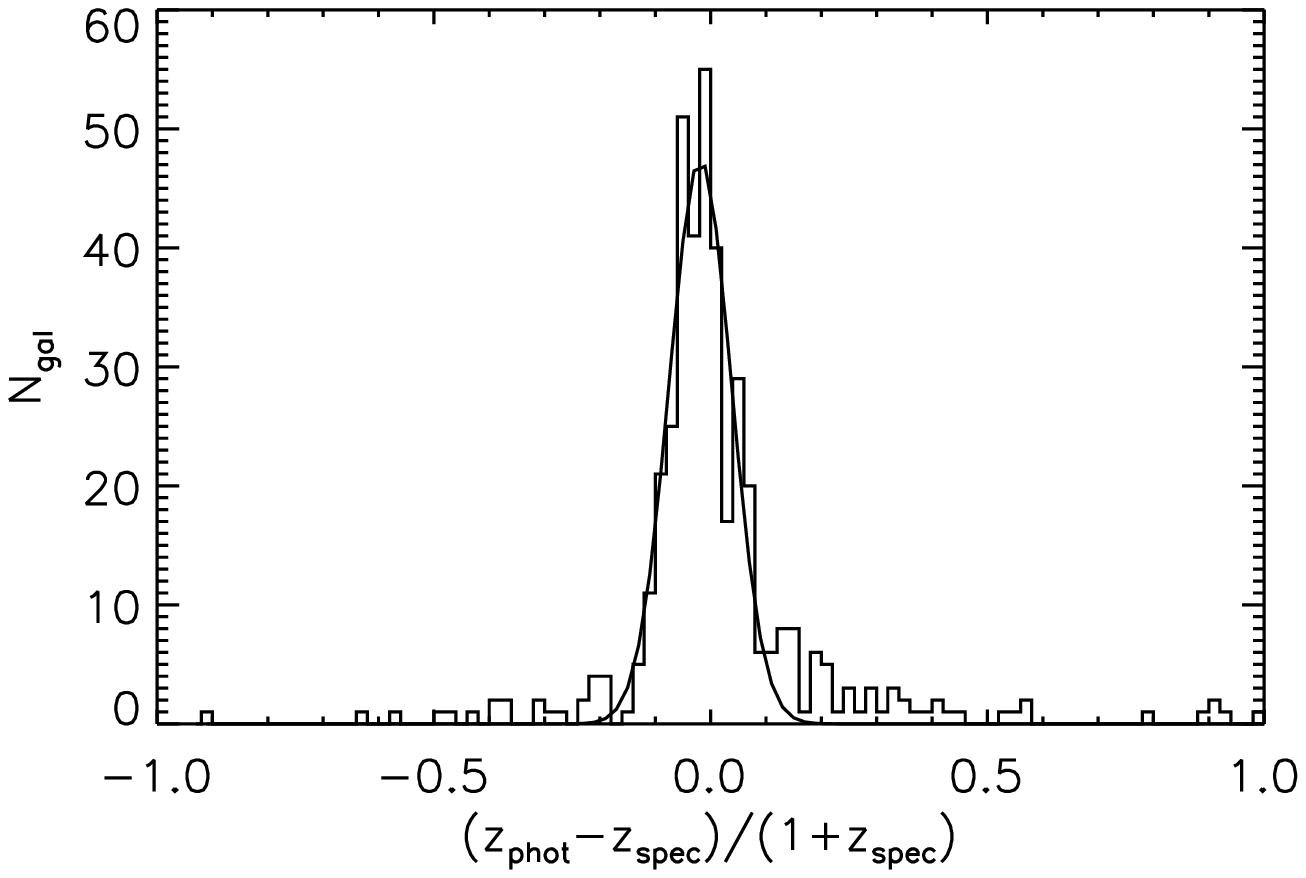}
\caption{\small Histogram of the difference between the
  photometric and spectroscopic redshifts. Overplotted is a Gaussian
  function fitted to the distribution. The mean difference $(z_{phot}
  - z_{\rm spec})/(1+z_{\rm spec})$ is -0.02. The error on the photometric
  redshift is $\sigma_z / (1+z_{\rm spec}) = 0.056$.}
\label{photospechist}
\end{figure}

We have matched our spectroscopic sample with our photometric redshift
catalogue (see VB06); the resulting diagram of $z_{\rm phot}$ versus
$z_{\rm spec}$ is shown in Fig.~\ref{photovsspec}. Overplotted is the
line for which $z_{\rm phot} = z_{\rm spec}$. It is apparent that most
objects lie along this line, however there are outliers, most of which
have photometric redshifts that are greatly overestimated. Closer
inspection of these objects reveals that the majority are AGN for
which our photometric redshift code is ill-suited as it does not
include the appropriate spectral energy distribution templates. A
histogram of the difference between the two redshift determinations,
scaled with redshift, is plotted in
Fig.~\ref{photospechist}. Overplotted is a Gaussian fit to the data;
the mean difference is $(z_{\rm phot} - z_{\rm spec})/(1+z_{\rm spec})
= -0.02$. The error on the photometric redshift is $\sigma_z /
(1+z_{\rm spec}) = 0.056$.

\begin{figure}
\includegraphics[width=90mm]{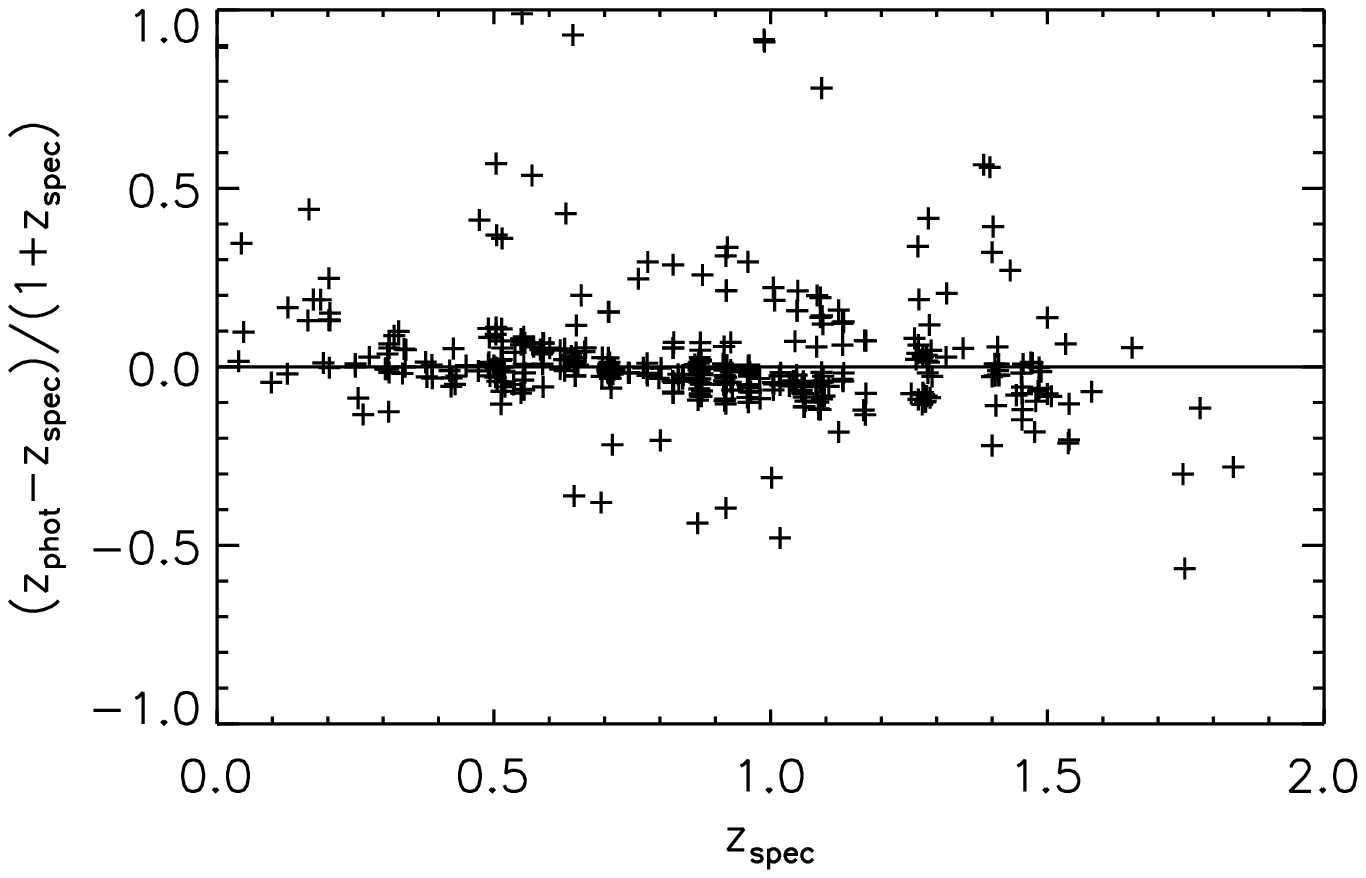}
\caption{\small The difference between the photometric and
  spectroscopic redshifts versus $z_{\rm spec}$}
\label{photospecdif}
\end{figure}

A more subtle effect seen in Fig.~\ref{photovsspec} is a `stepping' of
the photometric redshift along the $z_{\rm phot} = z_{\rm spec}$ line:
this is caused by the spikes in the photometric redshift
distribution. This is shown more clearly in Fig.~\ref{photospecdif};
here the difference between the two redshift determinations scaled
with redshift is plotted versus the spectroscopic redshift. It is
evident that the redshift spikes mainly comprise galaxies with
spectroscopic redshifts slightly deviating from the spike redshift, as
opposed to obvious outliers. This means that the redshifts of galaxies
just below the spike are slightly overestimated, and vice versa for
galaxies with slightly higher redshifts. This explains why the
redshifts of the clusters targeted with spectroscopy all seemed to be
underestimated by our algorithm (see Table~\ref{clust_z}), as they lie
at redshift $\sim 0.9$ which is just above the most prominent redshift
spike at $\sim 0.7$.

\newpage

\section{Tables of cluster members}

\begin{table*}
{\small
\begin{tabular}{lccrccc}
\hline
\noalign{\smallskip} 
\multicolumn{1}{c}{ID} & 
\multicolumn{1}{c}{RA} &  
\multicolumn{1}{c}{Dec.} &  
\multicolumn{1}{r}{$z_{\rm spec}$} & 
\multicolumn{1}{c}{$F_{\rm [OII]}$} &  
\multicolumn{1}{c}{$L_{\rm [OII]}$} &  
\multicolumn{1}{c}{$EW_0$} \\
\multicolumn{1}{c}{} &
\multicolumn{1}{c}{[h m   s]} &   
\multicolumn{1}{c}{[$^{\circ}$  $^{\prime}$  $^{\prime\prime}$]} &   
\multicolumn{1}{r}{} &
\multicolumn{1}{c}{[$10^{-20}~\rm W~m^{-2}$]} &
\multicolumn{1}{c}{[$10^{34}~\rm W$]} &
\multicolumn{1}{c}{[$\rm \AA$]} \\
\noalign{\smallskip} 
\hline 
\noalign{\smallskip} 
CVB6\_1   & 02:18:32.356   &  -05:00:51.23   &  0.86476 &    - &    -  &   -  \\ 
CVB6\_2   & 02:18:33.370   &  -05:01:03.87   &  0.86603 &    - &    -  &   -  \\ 
CVB6\_3   & 02:18:32.543   &  -05:01:27.26   &  0.86626 &    - &    -  &   -  \\ 
CVB6\_4   & 02:18:29.772   &  -04:59:42.89   &  0.86976 & 1.11 & 0.41  &  23  \\ 
CVB6\_5   & 02:18:34.818   &  -05:01:40.71   &  0.87001 &    - &    -  &   -  \\ 
CVB6\_6   & 02:18:35.391   &  -05:00:58.15   &  0.87052 &    - &    -  &   -  \\ 
CVB6\_7   & 02:18:32.239   &  -04:59:15.39   &  0.87056 & 1.50 & 0.56  &  5   \\ 
CVB6\_8   & 02:18:32.157   &  -04:59:24.70   &  0.87060 & 0.90 & 0.33  &  25  \\ 
CVB6\_9   & 02:18:37.447   &  -04:59:40.90   &  0.87064 & 1.59 & 0.59  &  42  \\ 
CVB6\_10  & 02:18:28.275   &  -05:00:05.84   &  0.87155 & 3.26 & 1.21  &  20  \\ 
CVB6\_11  & 02:18:35.286   &  -05:03:36.13   &  0.87162 & 2.87 & 1.07  &  5   \\ 
CVB6\_12  & 02:18:32.971   &  -05:00:51.11   &  0.87165 & 0.90 & 0.34  &  4   \\ 
CVB6\_13  & 02:18:29.653   &  -05:00:03.85   &  0.87180 & 1.45 & 0.54  &  25  \\ 
CVB6\_14  & 02:18:32.665   &  -04:59:24.59   &  0.87222 & 7.59 & 2.83  &  47  \\ 
CVB6\_15  & 02:18:32.789   &  -04:59:35.14   &  0.87294 & 3.57 & 1.33  &  36  \\ 
CVB6\_16  & 02:18:29.384   &  -05:01:24.56   &  0.87356 & 1.02 & 0.38  &  34  \\ 
CVB6\_17  & 02:18:41.821   &  -05:00:36.87   &  0.87499 & 7.40 & 2.78  &  54  \\ 
CVB6\_18  & 02:18:33.883   &  -04:59:41.42   &  0.87663 & 3.78 & 1.43  &  11  \\ 
CVB6\_19  & 02:18:38.941   &  -05:00:31.54   &  0.87753 &   -  &    -  &   -  \\ 
CVB6\_20  & 02:18:33.504   &  -05:01:03.57   &  0.87903 &   -  &    -  &   -  \\    
\noalign{\smallskip} 				    
\hline
\end{tabular}}  
\caption{\small
    Properties of the cluster galaxies of CVB6. Column 1 states the ID
    and the RA and Dec are given in columns 2 and 3. Column 4 is the
    heliocentric redshift; for the non-\oii~emitters this is measured
    via a cross-correlation technique with an estimated average error
    of $\sim 2 \times 10^{-4}$. For the \oii~emitters the redshift and
    line flux (column 5) are taken from a double Gaussian fit to the
    \oii~3727\,\AA\ line profile, with an average error of $\sim 5
    \times 10^{-5}$ in redshift, and $\sim 1 \times 10^{-21}\rm
    W\,m^{-2}$ in flux. The line luminosity is shown in
    column 6, and column 7 is the rest-frame equivalent width.}
\label{cvb6}
\end{table*}

\begin{table*}
\small{
\begin{tabular}{lccrccc}
\hline
\noalign{\smallskip} 
\multicolumn{1}{c}{ID} & 
\multicolumn{1}{c}{RA} &  
\multicolumn{1}{c}{Dec.} &  
\multicolumn{1}{r}{$z_{\rm spec}$} & 
\multicolumn{1}{c}{$F_{\rm [OII]}$} &  
\multicolumn{1}{c}{$L_{\rm [OII]}$} &  
\multicolumn{1}{c}{$EW_0$}  \\
\multicolumn{1}{c}{} &
\multicolumn{1}{c}{[h m   s]} &   
\multicolumn{1}{c}{[$^{\circ}$  $^{\prime}$  $^{\prime\prime}$]} &   
\multicolumn{1}{r}{} &
\multicolumn{1}{c}{[$10^{-20}~\rm W~m^{-2}$]} &
\multicolumn{1}{c}{[$10^{34}~\rm W$]} &
\multicolumn{1}{c}{[$\rm \AA$]} \\
\noalign{\smallskip} 
\hline 
\noalign{\smallskip} 
CVB11A\_1   & 02:18:12.399 &  -05:03:57.02  &   1.04765 &  1.31 &   0.77 &    26  \\ 
CVB11A\_2   & 02:18:11.004 &  -05:01:18.16  &   1.04799 &  2.99 &   1.75 &    31  \\ 
CVB11A\_3   & 02:18:20.645 &  -05:00:42.44  &   1.04897 &  1.83 &   1.08 &    80  \\ 
CVB11A\_4   & 02:18:04.326 &  -05:03:40.73  &   1.05744 &  0.90 &   0.54 &    14  \\ 
CVB11A\_5   & 02:18:04.760 &  -05:03:24.71  &   1.05865 &  4.03 &   2.42 &    49  \\ 
CVB11A\_6   & 02:18:05.479 &  -05:03:02.34  &   1.05874 &  8.98 &   5.40 &    63  \\ 
CVB11A\_7   & 02:17:57.969 &  -05:01:56.82  &   1.06013 &  3.53 &   2.13 &    57  \\ 
CVB11A\_8   & 02:18:42.363 &  -05:01:31.55  &   1.06311 &  11.9 &   7.27 &    69  \\ 
CVB11B\_9   & 02:18:27.599 &  -05:01:00.03  &   1.08248 &  1.88 &   1.20 &    29  \\ 
CVB11B\_1   & 02:18:27.599 &  -05:01:00.03  &   1.08295 &  1.11 &   0.71 &     9  \\ 
CVB11B\_2   & 02:18:02.998 &  -05:04:16.71  &   1.08416 &  1.44 &   0.92 &    19  \\ 
CVB11B\_3   & 02:18:02.509 &  -05:00:32.97  &   1.08440 &  1.12 &   0.72 &     3  \\ 
CVB11B\_4   & 02:17:57.227 &  -05:02:16.30  &   1.08795 &  5.85 &   3.77 &     2  \\ 
CVB11B\_5   & 02:18:01.459 &  -05:01:32.00  &   1.09003 &  5.10 &   3.30 &     9  \\ 
CVB11B\_6   & 02:18:00.503 &  -05:02:19.25  &   1.09231 &  1.64 &   1.07 &    11  \\ 
CVB11B\_7   & 02:18:24.556 &  -05:00:43.68  &   1.09276 &  3.83 &   2.49 &    33  \\ 
CVB11B\_8   & 02:18:01.169 &  -05:01:52.69  &   1.09441 &  2.73 &   1.78 &   137  \\ 
CVB11B\_9   & 02:17:55.399 &  -05:00:55.99  &   1.09472 &  4.51 &   2.95 &    56  \\ 
CVB11B\_10  & 02:17:52.096 &  -05:01:08.37  &   1.09559 &  1.38 &   0.90 &    12  \\ 
CVB11B\_11  & 02:17:53.995 &  -05:02:20.13  &   1.09785 &  3.34 &   2.20 &    66  \\ 
\noalign{\smallskip} 				    
\hline
\end{tabular}}   
\caption{\small Properties of the cluster galaxies of CVB11A and
  CVB11B. Columns are as in Table~\ref{cvb6}.}
\label{cvb11}
\end{table*}

\end{appendix}
\end{document}

%% file: table_targets.tex
\begin{tabular}{lcccrrrrrrr}
\hline
\noalign{\smallskip} 
\multicolumn{1}{c}{ID} & 
\multicolumn{1}{c}{$z_{\rm phot}$} &
\multicolumn{1}{c}{RA} &  
\multicolumn{1}{c}{Dec.} &  
\multicolumn{1}{c}{PA} &
\multicolumn{1}{r}{$N_{\rm P1}$} &
\multicolumn{1}{r}{$N_{\rm P2}$} &
\multicolumn{1}{r}{$N_{\rm P3}$} &
\multicolumn{1}{r}{$N_{\rm P1,MOS}$} &
\multicolumn{1}{r}{$N_{\rm P2,MOS}$} &
\multicolumn{1}{r}{$N_{\rm P3,MOS}$}\\
\multicolumn{1}{c}{} &
\multicolumn{1}{c}{} &
\multicolumn{1}{c}{[h m   s]} &   
\multicolumn{1}{c}{[$^{\circ}$  $^{\prime}$  $^{\prime\prime}$]} &   
\multicolumn{1}{c}{[$^{\circ}$]} &
\multicolumn{1}{r}{} &
\multicolumn{1}{r}{} &
\multicolumn{1}{r}{} &
\multicolumn{1}{r}{} &
\multicolumn{1}{r}{} &
\multicolumn{1}{r}{}\\
\noalign{\smallskip} 
\hline 
\noalign{\smallskip} 
CVB6    & 0.76  $\pm$ 0.12   &  02 18 32.7  &  -05 01 04  & 335 & 44 & 139 & 95 & 15 & 16 & 5 \\
CVB7    & 0.78  $\pm$ 0.06   &  02 19 03.5  &  -04 42 33  & 290 & 17 &  61 & 89 &  6 & 14 & 8 \\
CVB8    & 0.79  $\pm$ 0.07   &  02 17 54.0  &  -05 02 54  & 310 & 15 &  63 & 99 & 10 & 13 & 7 \\
CVB9    & 0.80  $\pm$ 0.06   &  02 17 21.4  &  -05 11 30  & 225 & 16 & 102 & 91 & 11 & 15 & 3 \\
CVB11   & 0.95  $\pm$ 0.11   &  02 18 06.7  &  -05 03 13  &  90 & 66 & 187 & 71 & 17 & 12 & 1 \\
\noalign{\smallskip} 
\hline    
\end{tabular}  